\documentclass[a4paper,11pt]{article}

\usepackage{amsmath,amssymb}           
\usepackage{amscd}                     
\usepackage{epsfig}                    
\usepackage[matrix,arrow]{xy}          
\usepackage{xspace}                    
\usepackage{stmaryrd}                  
\usepackage{slashed}
\usepackage{tabularx}

\usepackage{jheppub}                   
\makeatletter
\gdef\@fpheader{\ }                    
\makeatother

\newcommand{\dd}{\mathrm{d}}
\newcommand{\ee}{\mathrm{e}}
\newcommand{\ii}{\mathrm{i}}

\newcommand{\der}{\partial}

\newcommand{\bbZ}{\mathbb{Z}}
\newcommand{\bbR}{\mathbb{R}}
\newcommand{\bbC}{\mathbb{C}}

\newcommand{\CP}{{\bbC}P}

\DeclareMathOperator{\SU}{\mathit{SU}}
\DeclareMathOperator{\SO}{\mathit{SO}}

\DeclareMathOperator{\SL}{\mathit{SL}}
\DeclareMathOperator{\GL}{\mathit{GL}}

\DeclareMathOperator{\Spin}{\mathit{Spin}}
\DeclareMathOperator{\so}{\mathfrak{so}}
\DeclareMathOperator{\su}{\mathfrak{su}}

\DeclareMathOperator{\Cliff}{Cliff}

\newcommand{\rep}[1]{\mathbf{#1}}
\newcommand{\repp}[2]{(\rep{#1}, \rep{#2})}
\newcommand{\id}{\mathbf{1}}

\DeclareMathOperator{\tr}{tr}

\DeclareMathOperator{\vol}{vol}
\DeclareMathOperator{\Vol}{Vol}

\newcommand{\Lgen}{L}

\newcommand{\BLie}[2]{\left[#1,#2\right]}

\newcommand{\Dgen}{{D}}

\DeclareMathOperator{\AdS}{AdS}

\newcommand{\omegaB}{\omega_{\CP^3}}

\newcommand{\bl}[2]{\langle{#1},{#2}\rangle}

\newcommand{\proj}[1]{\times_{#1}}

\newcommand{\oadj}{\proj{\text{ad}}}

\DeclareMathOperator{\Edd}{\mathit{E_{d(d)}}}

\DeclareMathOperator{\E7}{\mathit{E}_{7(7)}}

\newcommand{\tA}{{\tilde{A}}}
\newcommand{\tF}{{\tilde{F}}}

\newcommand{\talpha}{{\tilde{\alpha}}}
\newcommand{\tbeta}{{\tilde{\beta}}}
\newcommand{\tLambda}{{\tilde{\Lambda}}}
\newcommand{\tV}{\tilde{V}}

\newcommand{\txi}{{\tilde{\xi}}}

\newcommand{\tC}{{\tilde{C}}}

\newcommand{\beq}{\begin{equation}}
\newcommand{\eeq}{\end{equation}}
\newcommand{\barr}{\begin{array}}
\newcommand{\earr}{\end{array}}

\newcommand{\hE}{\hat{E}}
\newcommand{\bE}{\bar{E}}

\newcommand{\bhE}{\bar{\hat{E}}}
\newcommand{\hF}{\hat{F}}

\newcommand{\gamh}{{\hat{\gamma}}}

\newcommand{\tpsi}{\tilde\psi}

\newcommand{\ba}{{\bar{a}}}
\newcommand{\bb}{{\bar{b}}}
\newcommand{\bc}{{\bar{c}}}

\newcommand{\dz}{\dd z}

\newcommand{\bw}{\bar{w}}
\newcommand{\dw}{\dd w}
\newcommand{\dwb}{\dd \bar{w}}

\newcommand{\bV}{\bar{V}}
\newcommand{\bW}{\bar{W}}
\newcommand{\bder}{\bar{\der}}
\newcommand{\tder}{\tilde{\der}}

\newcommand{\hs}[1]{\hspace{#1}}

\newcommand{\IIA}{\text{IIA}}
\newcommand{\Bbase}{b}
\newcommand{\Hbase}{h}

\title{New gaugings and non-geometry}

\author[a]{Kanghoon Lee,}
\emailAdd{kanghoon@kias.re.kr}
\author[b]{Charles Strickland-Constable}
\emailAdd{charles.strickland-constable@cea.fr}
\author[c,d]{and Daniel Waldram}
\emailAdd{d.waldram@imperial.ac.uk}

\affiliation[a]{Quantum Universe Center, 
   Korea Institute for Advanced Study, 
   Seoul 130-722, Korea} 

\affiliation[b]{Institut de physique th\'eorique, 
	Universit\'e Paris Saclay, CEA, CNRS, F-91191 Gif-sur-Yvette, France}

\affiliation[c]{Department of Physics,
   Imperial College London, 
   Prince Consort Road, London, SW7 2AZ, UK}

\affiliation[d]{Berkeley Center for Theoretical Physics,
   LeConte Hall MC 7300, \\
   University of California, Berkeley, CA 94720, U.S.A.}

\subheader{\textrm{Imperial/TP/15/DW/01}\\ \textrm{IPhT-T15/077}\\ \textrm{Q15003} \\ \textrm{UCB-PTH-15/01}}

\abstract{
We discuss the possible realisation in string/M theory of the recently discovered family of four-dimensional maximal $\SO(8)$ gauged supergravities, and of an analogous family of seven-dimensional half-maximal $\SO(4)$ gauged supergravities. We first prove a no-go theorem that neither class of gaugings can be realised via a compactification that is locally described by ten- or eleven-dimensional supergravity. In the language of Double Field Theory and its M theory analogue, this implies that the section condition must be violated. Introducing the minimal number of additional coordinates possible, we then show that the standard $S^3$ and $S^7$ compactifications of ten- and eleven-dimensional supergravity admit a new class of section-violating generalised frames with a generalised Lie derivative algebra that reproduces the embedding tensor of the $\SO(4)$ and $\SO(8)$ gaugings respectively. The physical meaning, if any, of these constructions is unclear. They highlight a number of the issues that arise when attempting to apply the formalism of Double Field Theory to non-toroidal backgrounds. Using a naive brane charge quantisation to determine the periodicities of the additional coordinates restricts the $\SO(4)$ gaugings to an infinite discrete set and excludes all the $\SO(8)$ gaugings other than the standard one.} 

\begin{document}
\maketitle


\section{Introduction}
\label{sec:intro}

It was only recently realised~\cite{Dall'Agata:2012bb} that the classic four-dimensional $\mathcal{N}=8$ gauged $\SO(8)$ supergravity theory of de Wit and Nicolai~\cite{deWit:1982ig} was actually one point in family of theories parametrised by an angle $0\leq\omega\leq\frac{1}{8}\pi$. The de Wit--Nicolai theory famously arises as a consistent truncation of eleven-dimensional supergravity on a round seven-sphere~\cite{dWN-S7,NP}. A natural question is whether all the theories in the family can be realised in higher-dimensional supergravity, or, more generally, as string or M theory backgrounds, a possibility that has been investigated by a number of authors~\cite{deWit:2013ija,GGN,Baron:2014yua,Dall'Agata:2014ita,Borghese:2014}\footnote{Very recently~\cite{Guarino:2015jca}, this question was also considered for the $ISO(7)$ gaugings of $N=8$ supergravity. Here, the family of gaugings consists of only two points~\cite{Dall'Agata:2014ita}: a purely electric gauging~\cite{Hull:1984vg} and a dyonic gauging. These arise as consistent truncations on $S^6$ of type IIA~\cite{Hull:1988jw} and massive type IIA supergravity respectively~\cite{Guarino:2015jca}.}.

In this paper, we first prove, using the formalism of $\E7\times\bbR^+$ generalised geometry~\cite{csw2,csw3,lsw1}, that only the de Wit--Nicolai theory can be realised in eleven dimensional supergravity. In fact the result is slightly stronger: the generic gaugings cannot be realised in any compactification that is locally described by eleven-dimensional or type II supergravity. Thus T- and U-fold geometries~\cite{Tfold}, that is backgrounds which are patched using T- or U-duality symmetries as for example in~\cite{Hellerman:2002ax,DH,Kachru:2002sk,Dabholkar:2005ve,Tfold1}, are also excluded. In the language of Double Field Theory (DFT)~\cite{dft} and its M theory analogue~\cite{BP}, it translates into the condition that the theories cannot be realised without violating the so-called ``strong constraint'' in DFT or the strong form of the ``section condition''~\cite{BP-alg,csw2} in the M theory analogue.
For Scherk-Schwarz reductions in DFT, it was shown in~\cite{relax2} that the weak form of the DFT constraint implies the strong form. A corresponding statement for extended geometries for the exceptional groups would then imply that the new gaugings must also violate the weak constraint. In the string theory case this corresponds to violating modular invariance. Such a no-go theorem is not an unexpected result, and a partial proof, applying only to $\SL(8,\bbR)$ subgroups of the full $\E7\times\bbR^+$ structure group, has already appeared in~\cite{Baron:2014yua}. 

In the second part of the paper we find a realisation of generic gaugings, using a minimal extension of conventional geometry, that explicitly violates the weak form of the section condition. We use the idea that the generalised Lie derivative algebra can still, in some cases, close on a set of frames that violate the weak constraint~\cite{relax1,relax2}. The new gaugings were originally constructed using the embedding tensor formalism~\cite{4D-gauge}. Finding an uplift translates geometrically into finding a generalised frame, which, under the generalised Lie derivative, reproduces the embedding tensor algebra. We showed in~\cite{lsw1} that the consistent truncations on the standard $S^7$ background considered by de Wit and Nicolai~\cite{dWN-S7} can be interpreted in generalised geometry as admitting precisely such a ``generalised Leibniz parallelisation''. As in conventional Riemannian geometry, a given background admits a family of different frames. In the case in hand these frames are related by local $\SU(8)/\bbZ_2$ rotations. The new result here is that the \emph{same} standard $S^7$ geometry admits a family of section-violating generalised frames that, using the standard expressions for the generalised Lie derivative~\cite{csw2}, reproduce the algebra of the full family of $\SO(8)$ gauge theories.

There is a family of $\SO(4)$ gaugings in seven-dimensional
half-maximal supergravity~\cite{DLR,DFMR} that closely parallels the
$\SO(8)$ gaugings of~\cite{Dall'Agata:2012bb}. A limiting case is the standard $\SO(4)$ gauging that arises from an $S^3$ compactification of $N=1$ ten-dimensional supergravity and corresponds to the near-horizon limit of multiple NS fivebranes. Again, as we show in appendix~\ref{app:nogo}, there is a no-go theorem that the generic gaugings cannot be lifted to any locally geometric background. The strong constraint of DFT must again be violated, and, following~\cite{relax2}, hence also the weak constraint. An earlier proof, using a slightly different notion of geometrical, was given in~\cite{DFMR}. The authors
of~\cite{DFMR} further give an explicit frame, violating the weak constraint, that reproduces the embedding tensor algebra of the generic gaugings. In this paper, we give an alternative realisation, showing that, in analogy with the $S^7$ case, the conventional $S^3$ background actually admits a family of generalised frames that violate the weak constraint but reproduce the embedding tensor algebra.

The $\SO(4)$ example is useful as it highlights several of the issues that arise in the $\SO(8)$ case in a much simpler purely string theory context. In particular, the standard $S^3$ background is just an $\SU(2)$ WZW model. Doubled geometries based on group manifolds have been discussed in~\cite{HR-E2,DPST}, and the relation betweens these models, DFT and T-folds in the case of $\SU(2)$ has been discussed very clearly in~\cite{Schulz:2011ye}. More recently a new version of DFT was derived for WZW backgrounds~\cite{Blumenhagen:2014gva}, giving an alternative description of the new gaugings. This connects to perhaps the most straightforward approach which is to find an explicit string conformal field theory description of the gaugings. The natural candidate suggested in~\cite{Blumenhagen:2014gva} is an ``asymmetric'' $\SU(2)$ WZW model with different levels on left and right. However this does not admit a modular-invariant partition function~\cite{Gannon:1992np}, reflecting a violation of the weak constraint.   

By definition, considering a frame that violates the section condition means that we are considering an ``extended spacetime'' with additional coordinates, as in DFT. For example, in the full M-theory version one would, in the context of $\SO(8)$ gaugings, replace the internal seven-dimensional supergravity manifold with some 56-dimensional space. There is a long history of introducing such extended spaces, and some reviews and references are given in~\cite{Aldazabal:2013sca,Berman:2013eva,Hohm:2013bwa,Hull:2014mxa}. This extended spacetime picture is in constrast to generalised geometry, where one considers structures on an extended generalised tangent space but the underlying manifold remains the conventional seven-dimensional one. 

The physical interpretation, if any, of these extra coordinates for generic backgrounds remains an open question. DFT began as a theory describing $d$-dimensional toroidal string backgrounds, where the extended space $\mathcal{X}$ is $2d$-dimensional space with a global flat $O(d,d)$ metric, and the extra coordinates come from the winding states of the string~\cite{dft}. Locally the section condition implies that all the fields are independent of half the coordinates and formally the field equations then reduce to those of the NSNS sector of type II supergravity in $d$ dimensions. However, globally one can also have more interesting configurations of the T-fold type, and it is the realisation of such exotic backgrounds, along with the torus T-duality, within an extended geometry that was the main motivation for the formulation. Subsequently the equations of DFT have been applied to much more general configurations, irrespective of whether any corresponding string winding modes exist. In addition, as already mentioned, study of WZW backgrounds suggests that the equations of DFT, including the section condition, may be more involved than those that appear from toroidal backgrounds~\cite{Blumenhagen:2014gva}, such that DFT is potentially background dependent.  

A number of questions naturally arise when attempting to extend DFT and its M-theory version to non-toroidal backgrounds~\cite{Hohm:2012gk,Park:2013mpa,Berman:2014jba,Cederwall:2014kxa,Papadopoulos:2014mxa,Hull:2014mxa}, most fundamentally perhaps, what the additional coordinates correspond to if there are no winding states. The fact that under the strong constraint the equations are formally equivalent to generalised geometry and hence supergravity suggests that, in this limit at least, DFT gives a valid description. However, even in this case the situation is actually more subtle. In generalised geometry the $O(d,d)$ metric and generalised Lie derivative exist on an extended tangent space over a conventional $d$-dimensional manifold. In DFT, one must extend these structures over a suitable doubled space $\mathcal{X}$ and this is generically problematic. The condition that all the fields are independent of half the coordinates defines a foliation, such that the $d$-dimensional conventional spacetime $M$ appears as the quotient of $\mathcal{X}$ by the action on the leaves (c.f.~\cite{Park:2013mpa}). In addition, the doubled space is typically required to admit a flat $O(d,d)$ metric. The existence of the metric and foliation structure is quite restrictive, and implies the existence of certain additional structures on $M$. Even dropping the requirement that the metric is flat does not allow one to describe generic supergravity backgrounds~\cite{Cederwall:2014kxa}. This has led to the suggestion that the doubled space is in some sense not a manifold~\cite{Hohm:2012gk}, perhaps admitting some non-associative structure~\cite{Hohm:2013bwa}. As noted in~\cite{Hull:2014mxa}, a key problem is that there is no embedding of the algebra of the generalised Lie derivative into the conventional Lie derivative algebra on the doubled space~$\mathcal{X}$. An alternative interpretation~\cite{Hull:2014mxa} is that, patch by patch, the $O(d,d)$ metric and generalised Lie derivative are defined not on $\mathcal{X}$ but on the $d$-dimensional quotient space as in generalised geometry.   

In addition, violating the weak constraint appears to be a radical step, since, at least in the original derivation for toroidal backgrounds, it implies violating level matching, and hence modular invariance. Generally the strong constraint is a necessary condition for the generalised Lie derivative to reproduce the closed algebra of diffeomorphisms and form-field gauge transformation of supergravity, the so-called generalised diffeomorphisms. Following~\cite{relax1,relax2}, it is possible to require only that a fixed generalised frame exists, giving a closed algebra under the generalised Lie derivative, in which case examples violating the strong (and weak) constraint can be found, and this is essentially the approach we adopt. However, it is important to note that in this case, only a finite dimensional algebra of symmetries survives, and this is not a subalgebra of some larger infinite dimensional space of generalised diffeomorphisms. As such the conditions for closure are considerably weaker. 

As should be clear from this discussion, finding an uplift of gauged supergravity to some extended spacetime theory is very far from providing a proof that the model can be realised in string or M theory. Our own position is that we remain sceptical about the physicalness of the extended coordinates of DFT and its M theory cousin beyond toroidal backgrounds, even with the section condition satisfied. That said, here we will attempt to give at least an internally consistent, minimalist approach to defining the extended geometry. We do not introduce the full set of doubled or extended coordinates, but only a minimum number, required to realise the algebra and such that the relevant $O(3,3)$ or $\E7$ structures, along with the generalised Lie derivative, can be explicitly defined in a covariant and global way. Thus our discussion can partly be viewed as an exercise in seeing what additional structures must be present when requiring the extended space to exist. Whether these descriptions are in any way physical remains an open question. 

The situation is clearest for the $S^3$ example, where, following~\cite{Schulz:2011ye}, we treat the extended space as T-fold, with the extra coordinate dual to strings wrapping the Hopf fibre, giving us confidence in the validity of the extended geometry. However the issue of the physical meaning of violating the strong (and weak) constraint remains. For the $S^7$ case the physical interpretation of the extra coordinates, while potentially related to wrapped D0-, D2-, D4- and D6-branes under the reduction to the type IIA on $\CP^3$, is actually considerably less clear. This is because the moduli spaces of the wrapped branes, other than for D0-branes, are not the geometrical $\CP^3$ and, unlike the T-fold case, there are no duality symmetries exchanging the different types of brane. In both cases, the brane interpretation fixes the periodicities of the extra coordinates and hence the global structure of the extended space. Interestingly, we then find that only a discrete, infinite, set of the non-geometric frames in $S^3$ case are globally defined (some of these also survive the quotient to the Lens space $S^3/\bbZ_p$). For the $S^7$ cases only the original geometrical frame is global, however, as mentioned, the brane interpretation of the extended geometry is much less well motivated in this case. 

The structure of the paper is as follows. Section~\ref{sec:strong-violation} discusses how consistent truncations are realised in generalised geometry and derives the no-go theorem for the $\SO(8)$ gaugings. The corresponding derivation for the $\SO(4)$ gaugings is given in appendix~\ref{app:nogo}. Section~\ref{sec:dft} includes a short discussion of some of the key aspects of extended geometries and the section condition, and the issues that arise when attempting to apply the formalism to non-toroidal backgrounds. Section~\ref{sec:s3} then defines the extended T-fold geometry for $S^3$, introduces the non-geometric frame and derives the corresponding gauge algebra. Section~\ref{sec:s7} defines the extended geometry and non-geometric frame for the $S^7$ background, and discusses the global properties. Much of the detailed calculation is relegated to the appendices. We conclude in section~\ref{sec:con}.


\section{A no-go theorem for $\SO(8)$ gaugings}
\label{sec:strong-violation}

We start by using generalised geometry to prove a no-go theorem: that
the new $\SO(8)$ gaugings cannot be realised as a compactification of a higher-dimensional theory that is locally geometrical. This means that, not only can they not can be realised as compactifications of eleven-dimensional supergravity, but they are also not described by the standard class of non-geometrical backgrounds. This is the class of spaces that are locally described by eleven-dimensional supergravity, but have no global geometrical description, such as U-folds and asymmetric orbifolds. Since we will consider them as a simpler toy model in what follows, for completeness, we also prove the equivalent no-go theorem for the family of $\SO(4)$ gaugings of half-maximal seven-dimensional supergravity in appendix~\ref{app:nogo}. 

Here we will focus only on supergravity, but, as we review in the next section, in the extended spacetimes models of DFT and its $\Edd$ cousin, it is standard to impose a ``section condition'' or ``strong constraint'' which restricts the dependence on the coordinates of the extended spacetime and is necessary to guarantee the closure of the gauge algebra. Under this condition, locally the theory is identical to conventional supergravity and hence we can again use a (local) generalised geometry analysis. Thus in the language of extended spacetimes, our no-go theorem implies that the new gaugings can only be realised if one violates the section condition. In~\cite{relax2} it was shown that, for DFT uplifts of gauged supergravity, if the frame violates the strong form of the section condition then it also violates the weak form, and one expects the same to hold for the $\Edd$ case. Since, in Double Field Theory, the weak constraint corresponds to level-matching in the string conformal field theory, dropping this condition is a significant and a priori hard to justify step. 

Let us start by recalling the form of the family of inequivalent $\SO(8)$ gaugings discovered in~\cite{Dall'Agata:2012bb}. The structure of a given  four-dimensional gauged $N=8$ supergravity is completely specified by an embedding tensor $X_{AB}{}^C$ transforming in the $\rep{912}$ representation of $E_{7(7)}$~\cite{4D-gauge}. For the particular gaugings of interest in this section, it is most convenient to study this object under the decomposition under the $\SL(8,\bbR)$ subgroup
\begin{equation}
\label{eq:X-SL8}
   \rep{912} = \rep{36} + \rep{36}' + \rep{420} + \rep{420}' .
\end{equation}
The new $\SO(8)$ gaugings of~\cite{Dall'Agata:2012bb} have only the $\rep{36}$ and $\rep{36'}$ parts non-zero 
\begin{equation}
\label{eq:Dall'Agata-embedding-tensor}
   X^{ij} = - R^{-1} \sin\omega \,\delta^{ij} , \qquad 
   \tilde{X}_{ij} = R^{-1} \cos\omega \,\delta_{ij} .
\end{equation}
The space of inequivalent theories is parameterised by $\omega \in [0, \tfrac{\pi}{4})$, with the standard geometrical $S^7$ case corresponding to $\omega = 0$. 


\subsection{Consistent truncations and generalised frames}
\label{sec:gen-frame}


In~\cite{lsw1}, it was argued that all consistent truncations of ten- or eleven-dimensional supergravity preserving maximal supersymmetry correspond to what was called a generalised Leibniz parallelisation on the compactification space. This is a particular type of preferred frame on the generalised tangent space. In particular we showed that the conventional $S^7$ truncation of eleven-dimensional supergravity, giving the $\SO(8)$ gauging with $\omega=0$, admitted precisely such a parallelisation. The proof of the no-go theorem depends on showing that such a frame does not exist, even locally, for general $\omega$. 

Let us first recall how such a frame is defined. The new gaugings admit an $\SO(8)$ invariant AdS vacuum. A local geometrical lift of the vacuum, is some patch of eleven-dimensional spacetime of the form $\AdS_4\times M$ where $M$ is some open set diffeomorphic to a patch of $\bbR^7$. In addition we have a metric 
\begin{equation}
\label{eq:metric}
   \dd s^2 = \ee^{2\Delta}\dd s^2(\AdS_4) + \dd s^2(M) , 
\end{equation}
and, generically non-trivial four-form flux $F$ and seven-form flux $\tF$ on $M$, where $\tilde{F}$ is the eleven-dimensional dual of the usual four-form flux on $\AdS_4$. The $\mathcal{N}=8$ supersymmetry implies we can also identify eight independent Killing spinors on $M$. 

The $\E7\times\bbR^+$ (exceptional) generalised geometry on $M$ is defined on the generalised tangent space~\cite{chris,PW}
\begin{equation}
\label{eq:E7E}
\begin{aligned}
   E &\simeq TM \oplus \Lambda^2T^*M \oplus \Lambda^5T^*M
         \oplus (T^*M \otimes\Lambda^7T^*M) , \\
   V &= v + \omega + \sigma + \tau ,
\end{aligned}
\end{equation}
which transforms as the $\rep{56}_{\rep{1}}$ representation under
$E_{7(7)}\times\bbR^+$ action, where a scalar $\rep{1}_{\rep{k}}$ of
weight $k$ under $\bbR^+$ is a section of $(\det T^*M)^{k/2}$. A generalised frame $\{ \hE_A \}$ in this geometry therefore carries an index $A = 1, \dots, 56$. In addition, given $V,V'\in E$ there is a generalised Lie derivative~\cite{csw2} given by\footnote{The notation here follows~\cite{PW,csw2,csw3}.}
\begin{equation}
\label{eq:Lgen-e7}
\begin{aligned}
   \Lgen_V V' &= (V\cdot\der)V' - (\der \oadj V)V' \\
      &= \mathcal{L}_v v'
          + \left( \mathcal{L}_v \omega' - i_{v'} \dd\omega \right)
          + \left( \mathcal{L}_v \sigma' - i_{v'} \dd\sigma
          - \omega'\wedge\dd\omega \right)
       \\ & \qquad
          + \left( \mathcal{L}_v \tau'
             - j\sigma'\wedge\dd\omega
             - j\omega'\wedge\dd\sigma \right) ,
\end{aligned}
\end{equation}
wich captures diffeomorphisms together with the gauge
transformations of three-form and dual seven-form gauge fields. 

The generalised metric $G$ defines a postive-definite inner product on the space of generalised vectors that is invariant under the maximal
compact subgroup $H_7=\SU(8)/\bbZ_2$. It unifies all the bosonic degrees of
freedom on the internal space, along with the warp factor $\Delta$. One can define an orthonormal frame by requiring, as usual,
\begin{equation}
\label{eq:ortho}
   G(\hE_A,\hE_B) = \delta_{AB} . 
\end{equation}
By definition, different orthonormal frames are related by $\SU(8)$ transformations. One can make this structure more explicit by decomposing
\begin{equation}
\begin{aligned}
   \{ \hE_A \} &= \{ \hE_{\alpha\beta} \} \cup \{ \bar{\hE}^{\alpha\beta} \} , \\
   \rep{56} &= \rep{28} + \bar{\rep{28}} . 
\end{aligned}
\end{equation}
where $\alpha$ and $\beta$ are $\SU(8)$ indices. 
For what follows we will also need the decomposition into a pair of real representations under the subgroup $\SO(8)\subset\SU(8)/\bbZ_2$, giving
\begin{equation}
   \{\hat{E}_{\alpha\beta}\}
       = \{\hat{E}_{ij}\} \cup \{\hat{E}^{\prime ij} \} ,
\end{equation}
with 
\begin{equation}
\label{eq:SU8-basis}
\begin{aligned}
   \hat{E}_{\alpha\beta}
      &= -\tfrac{1}{32}\ii \gamh^{ij}_{\alpha\beta}
         \big( \hat{E}_{ij} - \ii \hat{E}'_{ij} \big) , \\
   \bar{\hat{E}}^{\alpha\beta}
      &= \tfrac{1}{32}\ii \gamh^{ij\,\alpha\beta}
         \big( \hat{E}_{ij} + \ii \hat{E}'_{ij} \big) ,
\end{aligned}
\end{equation}
where we are matching the conventions of~\cite{csw2,csw3}. Full details of the conventions for gamma matrices and generalised geometry in $\SU(8)$ indices can be found in appendix~\ref{app:geom-SU8-indices}. The orthogonality conditions~\eqref{eq:ortho} then read
\begin{equation}
\label{eq:GSL8}
\begin{aligned}
   G(\hat{E}_{ij},\hat{E}_{kl})
      &= \delta_{ik}\delta_{jl} - \delta_{il}\delta_{jk} , \\
   G(\hat{E}_{ij},\hat{E}^{\prime\, kl})
      &= 0 , \\
   G(\hat{E}^{\prime\, ij},\hat{E}^{\prime\, kl})
      &= \delta^{ik}\delta^{jl} - \delta^{il}\delta^{jk} .
\end{aligned}
\end{equation}

Lifting the four-dimensional supergravity means we specify a particular metric and flux on $M$, and hence determines a particular generalised metric $G$. The eight independent complex Killing spinors spinors define a basis for $\SU(8)$ and hence a preferred frame $\{\hE_{\alpha\beta}\}$ for $G$. Hence, 
\begin{quote}
   \emph{The lift of any $\mathcal{N}=8$ gauged supergravity defines a preferred generalised frame $\{\hE_A\}$ on $M$.}
\end{quote}
This frame is defined up to $\SU(8)/\bbZ_2$ $R$-symmetry rotations, that is transformations that are constant on $M$ but can depend on the four-dimensional space coordinates $x^\mu$. If the lift is geometrical, that is $M$ is a full seven-dimensional manifold not just some open patch, the preferred frame is globally defined and hence gives generalised parallelisation of $E$. 

In addition, the frame $\{\hE_A\}$ encodes the embedding tensor $X_{AB}{}^C$ of the gauged supergravity via the generalised Lie derivative~\cite{ABMN,relax1,csw2,relax2,other-GSS}
\begin{equation}
\label{eq:embed-def}
   \Lgen_{\hat{E}_A} \hat{E}_B = X_{AB}{}^C \hat{E}_C . 
\end{equation}
The coefficients $X_{AB}{}^C$ are constant on $M$. If $M$ is global, this defines a ``generalised Leibniz parallelisation'', which can be viewed as the generalised geometry analogue of a local group manifold. It was argued in~\cite{lsw1} that all maximally supersymmetric consistent truncations are of this form. The lifts of generic points in the gauged supergravity are simply given by $\E7$ rotations of the preferred frame, in analogy with conventional Scherk--Schwarz reductions, 
\begin{equation}
\label{eq:SS}
   \hE_A'(x)= U_A{}^B(x) \hE_B
\end{equation}
where $U_A{}^B\in\E7$ and $x^\mu$ are coordinates in four dimensions. The inverse of the corresponding $x$-dependent generalised metric is given by 
\begin{equation}
\label{eq:G'}
   G'^{MN}(x) = \delta^{AB}\hE'_A{}^M(x) \hE'_B{}^N(x)
      = H^{AB}(x)\hE_A{}^M \hE_B{}^N ,
\end{equation}
where $H^{CD}(x)=\delta^{AB} U_A{}^C(x) U_B{}^D(x)$. Since the $\{\hE_A\}$ is defined only up to an $x$-dependent $\SU(8)$ rotation, the $U_A{}^B$ really parametrise a coset $\E7/(\SU(8)/\bbZ_2)$. These are the scalar degrees of freedom of the four-dimensional gauged supergravity. Using the results of~\cite{EFT1} one can write similar expressions for the four-dimensional gauge fields in terms of $\hE_A$~\cite{Hohm:2014qga}.

Finally we note that, using the Leibniz property of the generalised Lie derivative, acting with $\Lgen_{\hE_A}$ on~\eqref{eq:ortho} gives
\begin{equation}
\label{eq:GKV}
   ( \Lgen_{\hE_A} G ) (\hE_C,\hE_C)
      = - G(\Lgen_{\hE_A}\hE_B,\hE_C) - G(\hE_B,\Lgen_{\hE_A}\hE_C)
      = - X_{ABC} - X_{ACB} ,
\end{equation}
where $X_{ABC}=X_{AB}{}^D\delta_{CD}$. Recall that, viewed as matrices $X_A$ with components $(X_A)_B{}^C=X_{AB}{}^C$, the embedding tensor generates the gauge group $\mathcal{G}\subset\E7$ of the gauged supergravity~\cite{4D-gauge}. If $\mathcal{G}\subseteq\SU(8)/\bbZ_2$ then the action of $X_A$ must preserve the metric $\delta_{AB}$, implying $X_{AB}{}^D\delta_{DC}+X_{AC}{}^D\delta_{BD}=0$ and hence $\Lgen_{\hE_A}G=0$. Thus in this case the $\hE_A$ are all generalised Killing vectors.


\subsection{Proof of the no-go theorem}
\label{sec:nogo}


The embedding tensor for the $\SO(8)$ gaugings~\eqref{eq:Dall'Agata-embedding-tensor} implies that we require a generalised frame satisfying 
\begin{equation}
\label{eq:DallAgata-alg-frame}
\begin{aligned}
   L_{\hat{E}_{ij}} \hat{E}_{kl}
      &= 2R^{-1} \cos \omega \big(\delta_{i[k}\hat{E}_{l]j}
          - \delta_{j[k}\hat{E}_{l]i} \big) , \\
   L_{\hat{E}_{ij}} \hat{E}'_{kl}
      &= 2R^{-1} \cos \omega\big(\delta_{i[k}\hat{E}'_{l]j}
          - \delta_{j[k}\hat{E}'_{l]i} \big) , \\
   L_{\hat{E}'_{ij}} \hat{E}_{kl}
      &= -2R^{-1} \sin \omega \big(\delta_{i[k}\hat{E}_{l]j}
          - \delta_{j[k}\hat{E}_{l]i} \big) , \\
   L_{\hat{E}'_{ij}} \hat{E}'_{kl}
      &= -2R^{-1} \sin\omega \big(\delta_{i[k}\hat{E}'_{l]j}
          - \delta_{j[k}\hat{E}'_{l]i} \big) ,
\end{aligned}
\end{equation}
or equivalently
\begin{equation}
\label{eq:alg-SU8}
\begin{aligned}
   \Lgen_{\hat{E}_{\alpha\beta}} \hat{E}_{\gamma\delta}
      &= -\tfrac{1}{8}\ii R^{-1} \ee^{\ii\omega}\Big(
          \tC_{\alpha\gamma}\hat{E}_{\delta\beta}
          - \tC_{\alpha\delta} \hat{E}_{\gamma\beta}
          - \tC_{\beta\gamma} \hat{E}_{\delta\alpha}
          + \tC_{\beta\delta} \hat{E}_{\gamma\alpha}
          \Big) , \\
   \Lgen_{\hat{E}_{\alpha\beta}} \bar{\hat{E}}^{\gamma\delta}
      &= -\tfrac{1}{8}\ii R^{-1} \ee^{\ii\omega}\Big(
      	\delta_\alpha^\gamma
      	\tC_{\beta \epsilon} \bar{\hat{E}}^{ \delta \epsilon}
          - \delta_\beta^\gamma
      	\tC_{\alpha \epsilon} \bar{\hat{E}}^{ \delta \epsilon}
          - \delta_\alpha^\delta
      	\tC_{\beta \epsilon} \bar{\hat{E}}^{\gamma \epsilon}
          + \delta_\beta^\delta
      	\tC_{\alpha \epsilon} \bar{\hat{E}}^{\gamma \epsilon} \Big) , \\
\end{aligned}
\end{equation}
where the other relations follow from complex conjugation, and $\tC_{\alpha\beta}$ is the $\SO(8)$ gamma-matrix transpose intertwiner, which can always be chosen to be the identity (see appendix~\ref{app:gammas}). In addition, it is easy to check directly from~\eqref{eq:GKV} that all the basis vectors are generalised Killing, that is 
\begin{equation}
\label{eq:GKV2}
   \Lgen_{\hE_{ij}} G = \Lgen_{\hE'_{ij}} G = 0 . 
\end{equation}

The key ingredient in the proof is the graded form of the generalised Lie derivative~\eqref{eq:Lgen-e7}. In particular the vector part of $\Lgen_VV'$ can only come from the vectors in $V$ and $V'$, namely it is given by Lie bracket $\mathcal{L}_vv'=\BLie{v}{v'}$. Let us focus first on the first line of~\eqref{eq:DallAgata-alg-frame}. Note that in this case $\Lgen_{\hE_{ij}}\hE_{kl}=-\Lgen_{\hE_{kl}}\hE_{ij}$ and the first line is simply the $\so(8)$ Lie algebra. If we can expand the frame into its components
\begin{equation}
   \hE_{ij} = u_{ij} + \omega_{ij} + \sigma_{ij} + \tau_{ij} , 
\end{equation}
the grading of the generalised Lie derivative implies that vector components $u_{ij}$ satisfy
\begin{equation}
   \BLie{u_{ij}}{u_{kl}}
      = 2R^{-1}\cos\omega \big(
         \delta_{i[k}u_{l]j} - \delta_{j[k} u_{l]i} \big) . 
\end{equation}
In addition, the condition~\eqref{eq:GKV2} implies that 
\begin{equation}
   \mathcal{L}_{u_{ij}} g = \mathcal{L}_{u_{ij}} F
       = \mathcal{L}_{u_{ij}} \tF = 0 , 
\end{equation}
and so $u_{ij}$ are Killing vectors, and in addition preserve the fluxes. A priori, some of the vectors $u_{ij}$ may be identically zero. To see that this is not the case, let $\mathfrak{g}$ be the $\so(8)$ Lie algebra generated by the $\hE_{ij}$. Define a subalgebra $\mathfrak{k}\subset\mathfrak{g}$ of those elements with vanishing vector components. Given the grading of the generalised Lie derivative we have 
\begin{equation}
   \Lgen_V K \in \mathfrak{k}, \qquad 
   \forall V\in \mathfrak{g}, K\in \mathfrak{k} .
\end{equation}
In other words $\mathfrak{k}$ is an ideal. But $\so(8)$ is simple, and so has no non-trivial ideals, and hence $\mathfrak{k}=0$. Thus none of the $u_{ij}$ can vanish. 

We have shown that the uplift geometry admits 28 Killing vectors generating an $\so(8)$ Lie algebra. This implies that the metric and gauge fields, and hence the generalised metric $G$, are locally those of the round $S^7$ with constant $\tF$ proportional to the volume form, and vanishing $F$. Since locally the $\AdS_4$ background must satisfy the eleven-dimensional equations of motion, we see that, geometrically the putative uplift is exactly the same as for the standard $\SO(8)$ theory with $\omega=0$.

This result can also be seen by considerations of supersymmetry. Like the de Wit-Nicolai theory, all of the new gauged $\SO(8)$ theories possess an $\AdS_4$ vacuum at the origin of the scalar manifold which preserves the full $N=8$ supersymmetry in four dimensions~\cite{Dall'Agata:2012bb}. Any locally geometric uplift of this vacuum must therefore preserve maximal supersymmetry from the eleven-dimensional point of view. However, the only maximally supersymmetric $\AdS_4$ solution of eleven-dimensional supergravity is the $\AdS_4 \times S^7$ solution~\cite{FigueroaO'Farrill:2002ft}, so this must be the local geometry of the uplift.

Returning to our analysis of the generalised Lie derivative algebra, using the analogous argument for $\hE'_{ij}$, the last line of~\eqref{eq:DallAgata-alg-frame} implies that corresponding vector components $u'_{ij}$ also give 28 Killing vectors. A priori, these could be any linear combination of the $u_{ij}$, however the middle lines of~\eqref{eq:DallAgata-alg-frame} imply that they only differ by a rescaling. In other words we can conclude that 
\begin{equation}
\begin{aligned}
   \hE_{ij} &= \cos\omega v_{ij} + \dots \\
   \hE'_{ij} &= -\sin\omega v_{ij} + \dots 
\end{aligned}
\end{equation}
where the Killing vectors $v_{ij}$ satisfy~\eqref{eq:Killing-vecs}
\begin{equation}
   \BLie{v_{ij}}{v_{kl}}
      = 2R^{-1} \big(
         \delta_{i[k}v_{l]j} - \delta_{j[k} v_{l]i} \big) . 
\end{equation}
In~\cite{lsw1} we showed that the round $S^7$ admitted a generalised Leibniz parallelisation of the form
\begin{equation}
\label{eq:E7para}
   \hat{F}_A = \begin{cases}
      \hat{F}_{ij} = v_{ij} + \sigma_{ij} + i_{v_{ij}} \tilde{A}
         & \\
      \hat{F}^{\prime\, ij} = \omega_{ij} + \tau_{ij}
           - j\tilde{A}\wedge \omega_{ij}
         & 
      \end{cases}
\end{equation}
where $\omega_{ij}$, $\sigma_{ij}$ and $\tau_{ij}$ are defined
in~\eqref{eq:Killing-vecs} and~\eqref{eq:ost-def}. This frame gave the
standard $\SO(8)$ gauging with $\omega=0$. If we rotate to $\SU(8)$
indices we have\footnote{Note that since $\SO(8)$ is an automorphism of the algebra~\eqref{eq:DallAgata-alg-frame} both $\{\hE_A\}$ and $\{\hF_A\}$ are only defined up to global $\SO(8)$ rotations. We have used this freedom to align the Killing vectors in each to be $v_{ij}$.}
\begin{equation}
\begin{aligned}
   \hF_{\alpha\beta} &= v_{\alpha\beta} + \dots , \\
   \hE_{\alpha\beta} &= \ee^{\ii\omega} v_{\alpha\beta} + \dots ,
\end{aligned}
\end{equation}
where $v_{\alpha \beta} = -\tfrac{1}{32}\ii \gamh^{ij}_{\alpha\beta} v_{ij}$ and $+\dots$ denotes the non-vector parts. We have shown that the generalised metric is the same in both cases, and so $\{\hE_A\}$ and $\{\hF_A\}$ must be related by an $\SU(8)/\bbZ_2$ transformation, that is, we must be able to find a local $U_\alpha{}^\beta \in \SU(8)$ such that  
\begin{equation}
   \hE_{\alpha\beta} 
      = U_\alpha{}^\gamma U_\beta{}^\delta \hF_{\gamma\delta}. 
\end{equation}
In particular, for there to be a local uplift of the generic gauging, we need $U_\alpha{}^\beta$ such that 
\begin{equation}
\label{eq:SU8-v-stab-eqn}
   U_\alpha{}^\gamma U_\beta{}^\delta v_{\gamma \delta} 
      = \ee^{\ii \omega} v_{\alpha \beta} .
\end{equation}
We can consider the $\SU(8)$ invariant tensor, symmetric on four vector indices, given by 
\begin{equation}
   \kappa^{mnpq} 
      = \tfrac{1}{8!}\epsilon^{\alpha_1\dots\alpha_8}
         v^m_{\alpha_1\alpha_2}v^n_{\alpha_3\alpha_4}
         v^p_{\alpha_5\alpha_6}v^q_{\alpha_7\alpha_8} , 
\end{equation}
where we have written the explicit $m,n,\ldots=1,\dots,7$ vector indices on $M$. It is relatively easy to see that this is non-zero\footnote{It is an $\SO(8)$ invariant so must be of the form $\lambda g^{(mn}g^{pq)}$. The scalar $\lambda$ is given by the square of the self-dual object $K_{\alpha_1\alpha_2\alpha_3\alpha_4}$ defined in~\cite{dWN-S7} and hence does not vanish.}. Forming this invariant from both sides of~\eqref{eq:SU8-v-stab-eqn}, we find it has no solution for $U\in\SU(8)$ unless $\ee^{\ii \omega}$ is a fourth root of unity. In particular, in the range $\omega \in [0, \tfrac{\pi}{4})$ the only solution is the trivial one at $\omega=0$. Thus we can conclude, as required, that only the standard $\omega=0$ gauging admits a locally geometrical uplift. 

Finally, one might ask if looking for an uplift to type II theories, in particular type IIB, might avoid this theorem. The local supergravity theory is again described by $\E7\times\bbR^+$ generalised geometry~\cite{csw2,csw3} but with $M$ six-dimensional and a different decomposition of the generalised tangent space, namely~\cite{chris}
\begin{equation}
\label{eq:genEII}
   E \simeq TM \oplus T^*M \oplus \Lambda^\pm T^*M \oplus \Lambda^5T^*M 
       \oplus \left(T^*M\otimes\Lambda^6T^*M\right) ,
\end{equation}
where $\pm$ refers to Type IIA and IIB. Exactly the same arguments applied above can be used to show that the vector components $u_{ij}$ of $\hE_{ij}$ are all non-zero and form an $\so(8)$ isometry algebra under the conventional Lie bracket. However, this possibility is excluded since the maximal number of isometries in $d=6$ dimensions is $\frac{1}{2}d(d+1)=21$. Thus it is impossible to realise any of the $\SO(8)$ gaugings as locally geometric uplifts to type II.


\section{Extended theories and the section condition}
\label{sec:dft}


Before turning to how we might realise the generic $\SO(8)$ gaugings in an extended theory, let us briefly summarise some of the key aspects of DFT and the section condition~\cite{dft} as the primary example of such an extension, and then also the corresponding conditions for the $\E7$ extended geometry relevant to M-theory. This will also allow us to discuss some of the issues that arise when attempting to apply the formalism of extended geometry to non-toroidal backgrounds (see for example~\cite{Hohm:2012gk,Park:2013mpa,Berman:2014jba,Cederwall:2014kxa,Papadopoulos:2014mxa,Hull:2014mxa}). 

In the standard formulation of DFT one considers a patch $\mathcal{U}$ of some $2d$-dimensional space $\mathcal{X}$ with coordinates $X^M=(x^m,y_m)$ and a flat $O(d,d)$ metric 
\begin{equation}
\label{eq:DFTeta}
   \dd s^2 = \eta_{MN} \dd X^M \dd X^N = \dd x^m \dd y_m . 
\end{equation}
This structure was originally derived from string field theory on a toroidal background. In that case, the coordinates $X^M$ are picked out by the compactification: the $x^m$ are flat coordinates on $T^d$, dual to the momentum modes of the string, and $y_m$ are flat coordinates on the T-dual torus dual to the winding modes. More generally the coordinates $X^M$ are determined on a given patch $\mathcal{U}$ up to transformations that preserve the form of the metric $\eta$, that is, constant $O(d,d)$ rotations and translations
\begin{equation}
   X^{\prime M} = O^M{}_N X^N + C^M . 
\end{equation}
If $\mathcal{X}=T^{2d}$ the corresponding global rotation isometries give the discrete T-duality group $O(d,d;\bbZ)$. By taking more general, quotients $\bbR^{2d}/\Gamma$ for some freely acting discrete subgroup $\Gamma$ of the rotations and translations, the construction also gives a geometrical description of non-geometric T-fold backgrounds, where the space is patched using T-dualities~\cite{Dabholkar:2005ve,dft}.

In DFT one then considers fields over $\mathcal{X}$, in particular a generalised metric $G_{MN}$ encoding the conventional metric and $B$-field. One also defines the generalised Lie derivative, given $V,W\in T\mathcal{X}$ by
\begin{equation}
\label{eq:dft-gld}
   \Lgen_V W^M = V^N \nabla_N W^M + (\nabla^MV^N-\nabla^NV^M)W_N ,
\end{equation}
where indices are raised and lowered using $\eta$ and $\nabla_M$ is the (flat) Levi--Civita connection. This expression is usually written using the partial derivative $\der_M$ not the connection $\nabla_M$, with the assumption that one is using the flat coordinates $X^M$. Here, as in~\cite{Cederwall:2014kxa}, we use the Levi--Civita form to stress the covariance of the expression. (Obviously in the preferred coordinates $X^M$, we have $\nabla_M=\der_M$ and the expressions agree.) Note that generically the generalised Lie derivative does not satisfy the Leibniz condition and hence does not define an algebra.  

The strong constraint, or section condition, states that, given any two physical fields $f$ and $g$ (which may be tensors) on $\mathcal{X}$ one has 
\begin{equation}
   \eta^{MN} \nabla_M \nabla_N f = \eta^{MN} \nabla_M \nabla_N g = 0 , 
   \qquad
   \eta^{MN} \nabla_M f \nabla_N g = 0 , 
\end{equation}
where again we use the Levi--Civita connection to make the expressions covariant. Under the weak constraint only the first condition holds. Note that if the product $fg$ also satisfies the weak constraint then it is easy to see that the weak constraint implies the strong constraint. Thus the distinction only applies when naive products of physical fields are not physical. For vectors $V$ and $W$ satisfying the strong constraint, the generalised Lie derivative $\Lgen_VW$ also satisfies the strong constraint. Furthermore it satisfies the Leibniz condition and hence defines an algebra. 

For generalised Scherk--Schwarz reductions an important result is that if the frame satisfies the weak constraint then it also satisfies the strong constraint~\cite{relax2}. The argument is as follows. The frame itself $\hE_A$ must be physical because the uplift of the supergravity gauge fields depends on $\hE_A$. Similarly, the $x$-dependent generalised metric $G'$ given in~\eqref{eq:G'}, which gives the uplift of the scalar fields, must also be physical. This means that both $\hE_A$ and the product $\hE_A\otimes\hE_B$ appearing in $G'$ must satisfy the weak constraint. But that then implies each $\hE_A$ satisfies the strong constraint. The argument is actually slightly subtler since $G'$ does not depend on an arbitrary product of frames, however, using the additional information of the Leibniz proporty of the generalised Lie derivative, one can indeed show that the weak form implies the strong form. In terms of our no-go theorems, this is implies that the generic gauged supergravities actually cannot be realised without violating the weak constraint. 

The strong section condition is usually interpreted to mean that one looks for the most general solution of these conditions, given a large number of fields $f,g,\dots$. Locally the fields are independent of half the coordinates. Put another way, it defines a sub-bundle $N\subset T\mathcal{X}$, such that, on a physical field $f$, we have $V^M\nabla_M f=0$ for all $V\in N$, and in addition $N$ is a maximal, null sub-space, that is it is $d$-dimensional and 
\begin{equation}
   \eta(V,W)=0, \qquad 
   \text{for all $V,W \in N$} . 
\end{equation}
(One such subspace is for example the set of vectors of the form $V=v_m (\der/\der y^m)$.) By definition (for example by acting on scalar physical fields $f$) it is clear that 
\begin{equation}
   \text{if $V,W\in N$ then $\BLie{V}{W}\in N$} 
   \qquad
   \Leftrightarrow
   \qquad
   \text{$N$ is a foliation} ,
\end{equation}
where $\BLie{V}{W}$ is the conventional Lie bracket on $\mathcal{X}$. 

Locally one can always choose coordinates so that the solution of the strong constraint sets all the fields to be independent of $y_m$. Taking the quotient of $\mathcal{X}$ by the action of $\der/\der y_m$ on the leaves of the foliation gives some patch $U$ of $\bbR^d$, and the physical fields such as the generalised metric, descend to fields on $U$. The nomenclature ``section condition'' is thus somewhat confusing: it does not define a $d$-dimensional subspace of $\mathcal{X}$ but rather allows one to take a quotient, a point stressed in~\cite{Park:2013mpa}. Furthermore, the DFT generalised Lie derivative reduces to the standard  generalised Lie derivative of generalised geometry on $U$, also known as the Dorfman derivative. Thus, locally, under the strong constraint, DFT reduces to generalised geometry. 

As we have mentioned, the physical interpretation of the doubled spacetime $\mathcal{X}$ for non-toroidal backgrounds remains an open question, primarily because there is generically no simple interpretation of the extra $y_m$ coordinates as dual to winding modes, nor indeed is there generically any action of T-duality. Since we will be interested in extended geometries based on spheres, in particular $S^3$ in the case of DFT, we must address these issues. For the moment, one can simply assume a doubled spacetime $\mathcal{X}$ exists. The standard requirements then seem to be that there is
\begin{align*}
   \text{(1)} & \quad \text{$O(d,d)$ metric $\eta$,}  & 
   \text{(2)} & \quad \text{null foliation $N\subset T\mathcal{X}$.}
\end{align*}
It is typically assumed $\eta$ has the form~\eqref{eq:DFTeta} and so is flat. By the Killing--Hopf theorem this implies $\mathcal{X}=\bbR^{2d}/\Gamma$ (assuming it is complete). In addition, for standard geometrical backgrounds the quotient along the leaves of the foliation must exist and give a conventional $d$-dimensional spacetime $M$. 

The key point is that the existence of $\eta$ and $N$ imply the existence of additional structures that can strongly restrict what spaces are allowed. One can try to overcome this by relaxing the condition that $\eta$ is flat as in~\cite{Cederwall:2014kxa}. An alternative interpretation, proposed in~\cite{Hull:2014mxa}, assumes that the foliation of $\mathcal{X}$ exists, but $\eta$ and the generalised Lie derivative are defined only on the (local) $d$-dimensional quotient space, as in generalised geometry. The most radical suggestion to overcome the problem is to insist on the primacy of the doubled structure, and drop the condition that $\mathcal{X}$ is a manifold~\cite{Hohm:2012gk} with conventional diffeomorphism symmetry, though exactly what object it is remains unclear. In the proposal of~\cite{Hohm:2012gk}, one still introduces local coordinates $X^M$ so $\mathcal{X}$ does in fact still admit a manifold structure. One then tries to map the group generated by the generalised Lie derivative under the strong constraint into the conventional diffeomorphism group and use this to define a new notion of tensor. Related proposals have appeared in~\cite{Park:2013mpa} and~\cite{Berman:2014jba}. However, this map is not a group homomorphism~\cite{Hull:2014mxa}, so it is unclear in what sense $\mathcal{X}$ is supposed to realise the group structure of the generalised Lie derivative. 

Assuming more conservatively that $\mathcal{X}$ is a manifold with the conventional notion of tensors, in order to see the constraints implied by the existence of $\eta$ and $N$, let us try and construct the doubled space $\mathcal{X}$ corresponding to a generic conventional spacetime $M$. Naively, since locally the equations of DFT reduce to those of conventional supergravity, or more precisely the generalised geometry reformulation of supergravity, it would appear one can view any supergravity background in DFT. However, this is not the case. Generalised geometry formulates the geometry on the generalised tangent space $E\simeq TM\oplus T^*M$ over a $d$-dimensional manifold $M$. Given that $M$ is supposed to be a quotient of the doubled space $\mathcal{X}$ by the action on the leaves, the doubled space $\mathcal{X}$ must be the total $2d$-dimensional space of the cotangent bundle $\mathcal{X}=T^*M$ (or some compact quotient thereof), with the foliation corresponding to the fibres. Locally we have coordinates $(x^m,y_m)$ on $\mathcal{X}$ where $x^m$ are coordinates on $M$ and $y_m\dd x^m$ is a one-form at the point $x\in M$. This space has the correct topology since restricting to the zero section $y_m=0$ we have  $T\mathcal{X}|_M\simeq TM\oplus T^*M$. However, there is generically no natural flat metric on $\mathcal{X}=T^*M$. The best one can do is to allow non-flat metrics, as first considered in~\cite{Cederwall:2014kxa}. We use the fibration structure to define
\begin{equation}
   \label{eq:eta-fibre}
   \eta = \dd x^m \left( \dd y_m + \Gamma^p{}_{mn}(x)\,y_p\dd x^n \right) , 
\end{equation}
where $\Gamma^p{}_{mn}(x)$ is some $\GL(n,\bbR)$ connection on $TM$. We see immediately that defining $\eta$, and hence also the generalised Lie derivative, requires addition information: specifically a choice of connection on $M$. 
If one additionally takes the standard assumption that the metric on $\mathcal{X}$ is flat, we see that a necessary condition is that $\Gamma^p{}_{mn}$ must be pure gauge, meaning we can choose coordinates $x^m$ such that it disappears. But this implies that $TM$ is trivial, at least up to some discrete holonomy. In other words the $d$-dimensional manifold $M$ must be parallelisable (or some discrete quotient of a parallelisable manifold). This is consistent with the original observation that the double space is only naturally defined for toroidal backgrounds. 

These restrictions are in marked constrast to the generalised geometry picture. There one only requires $\eta$ and the generalised Lie derivative to be defined on the zero section $y_m=0$. But from~\eqref{eq:eta-fibre} we see that $\eta|_M=\dd x^m\dd y_m$ independent of the connection $\Gamma^p{}_{mn}$ and hence there \emph{is} a natural $O(d,d)$ metric in generalised geometry, without the need for additional structure. The same holds for the generalised Lie derivative. Thus generalised geometry can be defined for any differentiable manifold $M$.  The problem in DFT is precisely how one extends the generalised geometry $O(d,d)$ metric away from the zero section. Note that this picture is nonetheless consistent with the proposal of Hull~\cite{Hull:2014mxa} where $\eta$ and the generalised Lie derivative are only required to exist on the quotient space, which here is simply $M$ itself.   

Let us briefly summarise how the analogous constructions and obstructions appear in the $\E7$ extended geometry. One now considers a 56-dimensional space $\mathcal{X}$ with coordinates $X^M$, that admits a constant $\E7$ structure defined by the symplectic and symmetric quartic invariants
\begin{equation}
\label{eq:E7-inv}
   \Omega = \tfrac{1}{2}\Omega_{MN} \dd X^M \wedge \dd X^N , 
   \qquad 
   Q = Q_{MNPQ} \dd X^M \otimes \dd X^N \otimes \dd X^P 
      \otimes \dd X^Q ,
\end{equation}
where $\Omega_{MN}$ and $Q_{MNPQ}$ are the standard invariants (see for example~\cite{Cremmer:1979up}), independent of $X$. The geometry is invariant under global transformations of the form 
\begin{equation}
   X^{\prime M} = A^M{}_N X^N + C^M . 
\end{equation}
where $A$ is a global $\E7$ rotation\footnote{More precisely one should consider $\Omega$ and $Q$ to be defined only up to scale so that the symmetry becomes $\E7\times\bbR^+$.}. One can define a fixed torsion-free connection $\nabla_M$ which is equal to $\der_M$ in the particular $X^M$ coordinates. Thus $\mathcal{X}$ is an affine manifold and so, as in the $O(d,d)$ case, if it is complete, then $\mathcal{X}=\bbR^{56}/\Gamma$, where $\Gamma$ is a freely acting discrete subgroup of $\E7$ together with translations. Given physical fields $f$ and $g$ the strong section condition is given by~\cite{csw2}
\begin{equation}
\label{eq:sectionE7}
   P_{MN}{}^{PQ} \nabla_P \nabla_Q f 
      = P_{MN}{}^{PQ} \nabla_P \nabla_Q q = 0 , \qquad 
   P_{MN}{}^{PQ} \nabla_P f \nabla_Q g = 0 ,
\end{equation}
where $P_{MN}{}^{PQ}$ projects onto the 133-dimensional adjoint representation. From the general analysis of~\cite{Strickland-Constable:2013xta} one sees that one also needs to impose~\cite{EFT1}
\begin{equation}
   \Omega^{-1 MN} \nabla_M f \nabla_N g = 0 .  
\end{equation}
In the weak form of the condition, only the first condition in~\eqref{eq:sectionE7} holds. As in DFT, following~\cite{relax2}, one expects that, for a Scherk--Schwarz uplift, requiring $\hE_A$ to satisfy the weak constraint implies that it also satisfies the strong constraint, essentially because the generalised metric $G'$ given by~\eqref{eq:G'} must be physical. The full proof should require use of the Leibniz property of the generalised Lie derivative. For the no-go theorem of the previous section, this would imply that the generic $\SO(8)$ gaugings can only be realised by violating the weak section condition. 

As for DFT, only certain seven-dimensional supergravity manifolds $M$ can be realised in the extended theory. One takes $\mathcal{X}$ to be the total space $\mathcal{X}\simeq \Lambda^2T^*M \oplus \Lambda^5T^*M\oplus (T^*M\otimes\Lambda^7T^*M)$, or a compact quotient thereof. Just as before, defining general $\E7$ invariants $\Omega$ and $Q$ on $T\mathcal{X}$ requires a choice of $GL(7)$ connection $\Gamma$ on $TM$. If $\Omega$ and $Q$ are flat, as in~\eqref{eq:E7-inv}, then a necessary (but not sufficient) condition is that $M$ must be parallelisable. Again this is in contrast to generalised geometry, for which the $\E7$ structure exists for arbitrary manifolds $M$ -- the obstruction in the extended theory is again how one extends this structure away from the zero section. 

In summary, we have seen that requiring the existence of a doubled or extended manifold $\mathcal{X}$ is quite constraining. At the very least, the relevant generalised structure and section condition imply the existence of additional structures on the conventional $d$-dimensional space, namely a choice of connection, and in the standard formalism it implies $\mathcal{X}$ has the form $\bbR^{2d}/\Gamma$ (or $\bbR^{56}/\Gamma$ for $\E7$) and that $M$ is (at least) parallelisable. (An exception is the interpretation of~\cite{Hull:2014mxa} which only requires the generalised structure to exist on the (local) quotient.) For these reasons, in the following, we will adopt a minimalist approach to defining the extended geometry, defining as little additional structure as possible. In particular, we will not introduce the full set of doubled or extended coordinates, but only a minimum number, such that the generalised Lie derivative, can be explicitly defined in a covariant and global way.


\section{$S^3$ and non-geometric $\SO(4)$ gaugings}
\label{sec:s3}


In this section we consider the possible string theory realisation of a family of $\SO(4)$ gaugings of half-maximal, $d=7$ supergravity~\cite{DLR,DFMR}, parameterised by an angle $0\leq\omega\leq\frac{1}{4}\pi$. The standard $\SO(4)$ gauging that appears as a consistent truncation of the ten-dimensional $N=1$ supergravity on $S^3$ corresponds to $\omega=\frac{1}{4}\pi$, and, as we show in appendix~\ref{app:nogo}, there is again a no-go theorem stating that the generic gaugings cannot be realised as a locally geometric uplift to the ten-dimensional theory. From~\cite{relax2}, this implies that any uplift must violate the weak constraint. (An earlier proof, using a slightly different notion of geometric, has been given in~\cite{DFMR}.) Thus this family provides a closely related, but simpler, version of the issues one has to address in realising the new $\SO(8)$ gaugings of~\cite{Dall'Agata:2012bb}. 

The standard $S^3$ reduction of radius $R$ is simply an $\su(2)$ WZW model at level $q=R^2$~\cite{CHS}, so the most straightforward question is to ask if there is a conformal field theory description of the generic gauging. A natural proposal, coming from the analysis of DFT on WZW backgrounds given in~\cite{Blumenhagen:2014gva}, is an  ``asymmetric'' $\su(2)$ WZW model, with different levels on the left and right, such that the ratio is related to the angle $\omega$. However, such models are known not to admit a modular invariant partition function~\cite{Gannon:1992np}. This matches with the no-go theorem, since the weak constraint of DFT, strictly derived only for toroidal backgrounds, corresponds to level matching and hence modular invariance. Notwithstanding this, the uplift question was first addressed using DFT in~\cite{DFMR}, where an explicit non-geometric generalised frame was constructing, violating the weak constraint, such that the generalised Lie derivative algebra reproduced the generic $\SO(4)$ embedding tensor. In this case, the $\omega=\frac{1}{4}\pi$ point was realised using a non-geometrical background rather than as $S^3$. 

In the following we will give an alternative realisation of the gaugings as a doubled geometry, though again violating the strong and weak constraints. In this case, the underlying geometry is always the same: a round $S^3$ with $H$-flux. This means that the corresponding generalised metric $G$ is perfectly geometric, and describes a standard supergravity background. However, we choose a generalised frame for $G$ that is non-geometrical, that is, depends on the doubled space in a way that violates the weak constraint. Thus the description is only very ``weakly'' non-geometrical. Furthermore, the enlarged geometries we define are, in a second sense, the weakest extension of conventional geometry one might consider: we only extend one coordinate on the sphere (not all three), so that the extended geometry is well-defined as a T-fold. 

Using the global structure of the T-fold, we find that the parameter $\omega$ can only take discrete values. We then show that this construction can be extended to give $\SO(4)$ frames on the Lens spaces $S^3/\bbZ_p$, such that one can realise a larger, but still discrete, range of $\omega$ values. Since this construction violates the weak constraint, we are sceptical that it is giving a true string theory uplift. Nonetheless, within the language of DFT it corresponds to a minimal relaxation of the standard rules, and is perhaps a useful exercise in exploring the difficulties in defining extended geometries for generic backgrounds that have been mentioned in the previous section. 

Let us briefly summarise the structure of the $\SO(4)$ gaugings and the $S^3$ uplift of the $\omega=\frac{1}{4}\pi$ case. The general embedding tensor~\cite{half-embed,embed-lecture} for half-maximal seven-dimensional gauged supergravity is a three-form $X_{ABC}$ where  $A=1,\dots 6$ labels an $\SO(3,3)$ vector index. Decomposing under an $\SO(3)\times\SO(3)$ subgroup, the $\SO(4)$ gaugings correspond to the non-zero components~\cite{DFMR}
\begin{equation}
\label{eq:SO4-nongeom-X}
   X_{abc} = 2R^{-1}\epsilon_{abc} , \qquad
   X_{\bar{a}\bar{b}\bar{c}} 
      = 2R^{-1}\cot\omega\,\epsilon_{\bar{a}\bar{b}\bar{c}} ,
\end{equation}
with $0\leq\omega\leq\frac{1}{4}\pi$ so that $1\leq\cot\omega\leq\infty$ and where we regard two embedding
tensors related by a global $O(6,6)$ rotation as equivalent\footnote{Note that the $R^{-1}$ factor can always be scaled out of these components by an $\SO(6,6)$ transformation, but in what follows it is convenient to retain it.}. The
standard $\SO(4)$ gauging arising from a consistent truncation on $S^3$ corresponds to $\omega=\frac{1}{4}\pi$ while the $\omega=0$ limit defines an $\SO(3)$ gauging. To construct such gaugings as local geometrical lifts, we need a generalised frame $\{\hat{E}_A\}=\{\hat{E}^R_a,\hat{E}^L_{\bar{a}}\}$, satisfying the algebra
\begin{equation}
\label{eq:so4}
\begin{aligned}
  \Lgen_{\hat{E}^R_a}\hat{E}^R_b 
     &= X _{abc} \hat{E}^R_c , & &&
  \Lgen_{\hat{E}^L_{\bar{a}}}\hat{E}^L_{\bar{b}} 
     &= X_{\bar{a}\bar{b}\bar{c}} \hat{E}^L_{\bar{c}} , & &&
  \Lgen_{\hat{E}^L_{\bar{a}}}\hat{E}^R_a &= \Lgen_{\hat{E}^L_{\ba}} \hat{E}^R_a 
     = 0 . 
\end{aligned}
\end{equation}
We see that we have a pair of $\SU(2)$ algebras with different normalisations for $\hE^R_a$ an $\hE^L_{\ba}$. If $G$ is the corresponding generalised metric, it is easy to see that this algebra implies that the $\hE_A$ are generalised Killing vectors, that is $\Lgen_{\hE_A}G=0$. 

For the special case of $\omega=\frac{1}{4}\pi$, we showed in~\cite{lsw1} that on $S^3$ such a frame can be concretely constructed out of left- and right-invariant vectors and one-forms, as we now summarise. The near-horizon limit of $q$ parallel NS fivebranes is described by a spacetime of the form $\bbR^{5,1}\times\bbR_t\times S^3$ with metric, $B$-field and dilaton given by~\cite{CHS}
\begin{equation}
\label{eq:S3pq-soln}
\begin{aligned}
   \dd s^2 &= \dd s^2(\bbR^{5,1}) + \dd t ^2 
       + \tfrac{1}{4}R^2\left[ \dd\theta^2 + \sin^2\theta \dd \phi^2
          + \left(\dd\psi + \cos\theta\dd\phi\right)^2 \right] , \\
   B &= \tfrac14 R^2 \cos\theta \dd\phi\wedge \dd\psi , \\
   \phi &= - t/R , 
\end{aligned}
\end{equation}
where Hopf fibre coordinate $\psi$ has periodicity $\psi\sim\psi+4\pi$ and we take units where $\alpha'=1$. 
Integrating $H$ over $S^3$ one sees that the quantisation of the $H$-flux gives $R^2=q\in\bbZ$. The solution preserves half of the supersymmetries of type II. T-dualising the Hopf fibre, one obtains the Lens space $S^3/\bbZ_q$ with one unit of $H$ flux~\cite{Maldacena:2001ky}. This solution has the same local form but with a Hopf fibre coordinate $\tpsi$ with periodicity $\tpsi\sim \tpsi+4\pi/q$. 

One can construct a global generalised frame on the generalised tangent space $E\simeq TS^3\oplus T^*S^3$ given by~\cite{lsw1},  
\begin{equation}
\label{eq:FLR}
\begin{aligned}
   \hF^L_a &= l_a - \lambda_a - i_{l_a}B , \\
   \hF^R_{\bar{a}} &= r_{\bar{a}} + \rho_{\bar{a}} 
      - i_{r_{\bar{a}}}B , 
\end{aligned}
\end{equation}
where $l_a$ and $r_a$ are the usual left- and right-invariant vectors
and $\lambda_a$ and $\rho_a$ the corresponding forms
given in~\eqref{eq:inv-vec} and~\eqref{eq:inv-form}. Explicitly
\begin{equation}
\label{eq:ER}
\begin{aligned}
   \hF^R_+ 
      &= \ee^{\ii\phi} \Big[ 
         \left( 2R^{-1}\der_\theta + \tfrac{1}{2}R\dd\theta \right)
         \\ & \qquad \qquad 
         + \ii\cot\theta \left( 2R^{-1}\der_\phi - \tfrac{1}{2}R\dd\phi \right)
         - \ii\csc\theta\left( 2R^{-1}\der_\psi + \tfrac{1}{2}R\dd\psi \right)
         \Big] , \\
   \hF^R_3 & 
      = 2R^{-1}\der_\phi + \tfrac{1}{2}R\dd\phi ,
\end{aligned}
\end{equation}
and similarly
\begin{equation}
\label{eq:EL}
\begin{aligned}
   \hF^L_+ 
      &= \ee^{-\ii\psi} \Big[ 
         \left( 2R^{-1}\der_\theta - \tfrac{1}{2}R\dd\theta \right)
         \\ & \qquad \qquad 
         + \ii\csc\theta \left( 2R^{-1}\der_\phi - \tfrac{1}{2}R\dd\phi \right)
         - \ii\cot\theta\left( 2R^{-1}\der_\psi 
               + \tfrac{1}{2}R\dd\psi \right) \Big] , \\
   \hF^L_3 & 
      = 2R^{-1}\der_\psi - \tfrac{1}{2}R\dd\psi , \\
\end{aligned}
\end{equation}
where $\hF^R_+=\hF^R_1+\ii\hF^R_2$ etc. The
algebra~\eqref{eq:embed-def} of this frame under the generalised Lie
derivative is precisely~\eqref{eq:so4} with $\omega=\frac{1}{4}\pi$. 


\subsection{The T-fold double geometry}
\label{sec:T-fold}

Our goal is to construct an extended geometry corresponding to $S^3$ with a global section-violating frame on $S^3$ that reproduces the general algebra~\eqref{eq:so4}. A clue to how we might do this is to recall that under T-duality along the $\psi$ Hopf fibration, the $S^3$ solution with $q$ units of flux is dual to the Lens space $S^3/\bbZ_q$ with one unit of flux. Consider the frame $\{\hF_A\}$ given in~\eqref{eq:FLR} on the Lens space, where we label the Hopf fibre coordinate $\tpsi$. We find that the $\hF^R_a$ frame is still globally defined, but not the $\hF^L_\ba$ frame since on $S^3/\bbZ_q$ the periodicity of the Hopf fibre is $\tpsi\sim\tpsi+4\pi/q$. Thus it appears that $S^3/\bbZ_q$ does not admit a global frame. Correspondingly there are fewer global Killing spinors and the $S^3/\bbZ_q$ background only preserves one-quarter of the type II supersymmetry. However, we still have the original frame defined on the dual $S^3$, depending not on $\tpsi$ but the dual $\psi$ coordinate, along with the corresponding Killing spinors. Thus if we allow the frame to depend on the dual $\psi$ coordinate, the $S^3/\bbZ_q$ still admits a global frame and preserves half the supersymmetries. This is the phenomenon of ``supersymmetry without supersymmetry'' as analysed for $S^5$ in~\cite{Duff:1998us}. 

This immediately suggests a generalisation where we allow the frame to depend on both $\psi$ and $\tpsi$. One starts by defining a ``T-fold''~\cite{Tfold}, a four-dimensional space $X$ composed of an $S^2$ base, parametrised by $\theta$ and $\phi$, with two circles fibered over it: the Hopf fibration of the original $S^3$ with coordinate $\psi$ and the dual circle fibration for $S^3/\bbZ_q$ with coordinate $\tpsi$, so 
\begin{equation}
\label{eq:X4}
   \begin{CD}
      S^1_\psi \times S^1_{\tpsi} @>>> X \\
      @. @V{\pi}VV \\
      @. S^2_{\theta,\phi}
   \end{CD}  
\end{equation}
Such geometry has been discussed in some detail in~\cite{Schulz:2011ye}. Note that there is no conserved winding number to generate the charge for the corresponding $S^1_{\tpsi}$ circle since the fundamental group $\pi_1(S^3)$ is trivial, though there are corresponding extended string states wrapping great circles. We take the periodicities 
\begin{equation}
\label{eq:psi-period}
   \psi \sim \psi + 4\pi, \qquad 
   \tpsi \sim \tpsi + 4\pi/q ,
\end{equation}
with $R^2=q$. The T-duality map identifies
\begin{equation}
\label{eq:Tdpsi}
   \tfrac{1}{2}R \dd\psi \simeq  2R^{-1}\der_{\tpsi}
\end{equation}
and similarly for $\frac{1}{2}R\dd\tpsi\simeq 2R^{-1}\der_\psi$. The space $X$ is commonly referred to as a ``correspondence space" (see e.g.~\cite{Belov:2007qj}), which possesses projections onto both the physcial background and the T-dual space. Although we focus on the $S^3$ case, the space $X$, and the extended geometry described below, can be defined in exactly the same way for more general $S^1$ fibration backgrounds.
 
One can clearly view the $S^1\times S^1$ fibre as a conventional double torus geometry with an $O(1,1)$ metric given by $\dd s^2= \tfrac14 R^2 \dd\psi\dd\tpsi$. However, we can also consider a sort of hybrid doubled geometry on $X$ itself. It is not conventional generalised geometry, which would only describe objects on $S^3$, nor is it the full double geometry which would require some six-dimensional space. Instead, it is the minimal extension allowed, introducing an explicit dual coordinate $\tpsi$ for a single isometry $\der_\psi$. The six-dimensional generalised tangent space on $X$ is built from generic vectors $v$ together with one-forms $\lambda$ that are sections of a sub-bundle pulled back from $S^2$, namely   
\begin{equation}
\label{eq:E-Tfold}
\begin{aligned}
   E &\simeq TX \oplus \pi^* T^*S^2 ,  \\
   V &= v + \lambda . 
\end{aligned}
\end{equation}
The identification~\eqref{eq:Tdpsi} means that the vector component $v^{\tpsi}$ should transform like the one-form $\lambda_\psi$ and similarly for $v^\psi$ and $\lambda_{\tpsi}$. This actually defines $E$ as a somewhat unconventional extension 
\begin{equation}
   \pi^*T^*S^2 \longrightarrow E \longrightarrow TX .
\end{equation}
We write 
\begin{equation}
   v = v^\alpha\der_\alpha + v^\psi \der_\psi + v^{\tpsi}\der_{\tpsi} , 
   \qquad
   \lambda = \lambda_\alpha \dd x^\alpha
\end{equation}
where $x^\alpha$ are coordinates on $S^2$. Suppose on the overlap $U_i\cap U_j$ of two patches of $S^2$, the $S^1$ fibres are patched by 
\begin{equation}
   \psi_{(i)} = \psi_{(j)} + \Lambda_{(ij)}, \qquad 
   \tpsi_{(i)} = \tpsi_{(j)} + \tLambda_{(ij)}, 
\end{equation}
then we have 
\begin{equation}
\label{eq:Tfold-patch}
\begin{aligned}
   v^\alpha_{(i)} &= v^\beta_{(j)} , \\
   v^\psi_{(i)} &= v^\psi_{(j)} + v^\alpha_{(j)}\der_\alpha\Lambda_{(ij)} , \\
   v^{\tpsi}_{(i)} 
      &= v^{\tpsi}_{(j)} + v^\alpha_{(j)}\der_\alpha\tLambda_{(ij)} , \\
   \lambda_{(i),\alpha} &= \lambda_{(j),\alpha}
   	+ v^\beta_{(j)} (\dd \hat{\Lambda}_{(ij)})_{\beta\alpha} 
       - \tfrac{1}{4}R^2 \Big[ v^\psi_{(j)}\der_\alpha\tLambda_{(ij)}
       + v^{\tpsi}_{(j)}\der_\alpha\Lambda_{(ij)} 
        \\ & \hspace{145pt} 
	+ \tfrac12 [ 
           (v^\beta\der_\beta\Lambda_{(ij)}) \der_\alpha\tLambda_{(ij)} 
           + (v^\beta\der_\beta\tLambda_{(ij)}) \der_\alpha \Lambda_{(ij)}] \Big] .
\end{aligned}
\end{equation}
Here $\hat{\Lambda}\in T^*S^2$ describes the patching of the $B$-field on the base $S^2$. Concretely, given $\Bbase_{(i)}\in \pi^*\Lambda^2T^*U_i$, we have 
\begin{equation}
\label{eq:Bbase-patching}
   \Bbase_{(i)} = \Bbase_{(j)} - \dd \hat\Lambda_{(ij)} 
      - \tfrac{1}{8}R^2 A_{(j)} \wedge \dd \tLambda_{(ij)} 
      - \tfrac{1}{8}R^2 \tA_{(j)} \wedge \dd \Lambda_{(ij)} .
\end{equation}
Crucially, there is an $O(3,3)$ metric on $E$ given by 
\begin{equation}
\label{eq:dft-eta}
\begin{aligned}
   \eta(V,V) 
      & = v^\alpha\lambda_\alpha
           + \tfrac{1}{4}R^2v^\psi v^{\tpsi} .
\end{aligned}
\end{equation}

As stands this form of the metric does not look covariant. However, this is simply due to the unconventional patching~\eqref{eq:Tfold-patch}, which implies that $\lambda=\lambda_\alpha\dd x^\alpha$ is not globally a one-form on $X$. We would like to show explicitly that $\eta$ is indeed covariant and also to stress what additional structures on $X$ are required to define the extended geometry. This is most naturally done, by defining the ``untwisted'' one-forms  
\begin{equation}
\label{eq:untwist}
   \tilde{\lambda}_\alpha 
     = \lambda_\alpha + v^{\beta} \Bbase_{\beta \alpha} 
     	- \tfrac{1}{4}R^2  \Big[ A_\alpha v^{\tpsi} + \tA_\alpha v^\psi 
        + \tfrac{1}{2}(v^\beta A_\beta)\tA_\alpha
              + \tfrac{1}{2} (v^\beta \tA_\beta)A_\alpha \Big] ,
\end{equation}
which are true sections of $\pi^*T^*S^2\subset T^*X$. The $O(3,3)$ metric~\eqref{eq:dft-eta} then takes the manifestly covariant form 
\begin{equation}
\label{eq:eta-T-fold}
   \eta(V,V) = i_v\tilde{\lambda} + \tau(v,v) , 
\end{equation}
where  
\begin{equation}
   \tau = \tfrac{1}{8}R^2 \Big[ 
        (\dd\psi + A)\otimes (\dd\tpsi + \tA)
        + (\dd\tpsi + \tA)\otimes (\dd\psi + A) \Big]
\end{equation}
is an $O(1,1)$ metric on the $T^2$ fibres. We see explicitly that defining $\eta$ requires additional structure beyond just the topology of $X$. In particular, it requires knowledge of the fibration structure defined by $\pi$ and the metric on the fibres $\tau$. This matches our discussion in section~\ref{sec:dft}: the existence of an $O(d,d)$ metric actually puts additional structure on the doubled space. 

Returning to the original $S^3$ geometry corresponds to ``solving the section condition'' by taking a quotient by the $U(1)$ symmetry generated by $\der_{\tpsi}$ and using the metric $\tau$ to identify the vector $v^{\tpsi}\der_{\tpsi}$ with the one-form $\frac{1}{4}R^2v^{\tpsi}\dd\psi$. Note that, despite the name, this is a quotient not a choice of section: the $S^1_{\tpsi}$ bundle is non-trivial and so does not admit a global section. Alternatively one can solve the section condition by taking a quotient by $\der_\psi$ giving the dual $S^3/\bbZ_p$ geometry. After such quotients the connections $\tA$ and $A$ becomes the $B$-field component $B_i=\dd\psi\wedge\tA_i+\Bbase_i$ and $\tilde{B}_i=\dd\tpsi\wedge A_i+\Bbase_i$ respectively, as is usual in T-duality\footnote{Note that we have included the $\Bbase$ components in order to preserve the full gauge invariance of the geometry. In the $S^3$ example we can always choose a gauge such that e.g. $B = \dd \psi \wedge \tA$ after taking the $\der_{\tpsi}$ quotient. The patching~\eqref{eq:Bbase-patching} of $\Bbase$ includes Chern-Simons terms for the connections $A$ and $\tA$ as is standard in the $S^1$ reduction of generalised geometry (see for example~\cite{Coimbra:2014qaa}).}.

Finally, we can also define a notion of generalised Lie derivative that reduces to the conventional generalised geometry expressions on $S^3$ and $S^3/\bbZ_p$ after reduction along $\der_\psi$ or $\der_{\tpsi}$ respectively. In the particular coordinates where $V^M=(v^\theta,v^\phi,v^\psi,v^{\tpsi},\lambda_\theta,\lambda_\phi)$, we have the standard DFT form as in~\eqref{eq:dft-gld}, 
\begin{equation}
\label{eq:Lgen-coord}
   \Lgen_V W^M = V^N \der_N W^M + \left(\der^MV^N-\der^NV^M\right)W_N ,
\end{equation}
where indices are raised and lowered using the metric~\eqref{eq:dft-eta} and the partial derivative is given by $\der_M=(\der_\theta,\der_\phi,\der_\psi,\der_{\tpsi},0,0)$. As discussed in the previous section, this is not a covariant expression, so cannot be used as a generic definition of $\Lgen_VW$ without specifying how the coordinates are chosen.

However, we can write an intrinsically covariant expression for $\Lgen_V W$ in terms of the untwisted objects $V=v+\tilde\lambda$ and $W=w+\tilde\mu$. We have
\begin{equation}
\label{eq:GLDu}
\begin{aligned}
   \Lgen_V W &= \BLie{v}{w} + \tfrac{8}{R^2}\big[ \eta(\mathcal{L}_\xi V,W)\txi
       + \eta(\mathcal{L}_\txi V,W)\xi \big]
       \\ & \qquad
       + \left[ \mathcal{L}_v \tilde\mu - i_w\dd \tilde\lambda 
          + \tfrac{16}{R^2}\big[  \eta(W,\txi)\dd \eta(V,\xi) 
          + \eta(W,\xi)\dd \eta(V,\txi) \big]
          \right]_{B} \\
         & \qquad  + i_v i_w \Hbase + \tfrac{1}{4}R^2 \big[ (i_w \sigma) (i_v \tF) - (i_v \sigma) (i_w \tF) + (i_w \tilde{\sigma}) (i_v F) - (i_v \tilde{\sigma}) (i_w F) \big]
\end{aligned}
\end{equation}
where 
\begin{equation}
   \xi = \der_\psi  \qquad \text{and} \qquad \txi = \der_{\tpsi} , 
\end{equation}
are the two vector fields defined by the $U(1)$ actions on $S^1_\psi$ and $S^1_{\tpsi}$, 
\begin{equation}
   \sigma = \dd \psi + A \qquad \text{and} \qquad \tilde{\sigma} = \dd \tpsi + \tA,
\end{equation}
are the dual one-forms in the metric~\eqref{eq:eta-T-fold} and
\begin{equation}
   [\alpha]_B = \alpha - (i_\xi\alpha) \sigma
       - (i_{\txi}\alpha) \tilde{\sigma}
\end{equation}
is the projection of $\alpha\in T^*X$ onto $\pi^*T^*S^2$.
The additional flux terms in~\eqref{eq:GLDu} are typical of generalised Lie derivatives expressed in terms of untwisted variables and feature the $U(1)$ curvatures of the fibered cirlces $F = \dd A$ and $\tF = \dd \tA$ as well as the gauge-invariant field strength for the two-form $\Bbase$,
\begin{equation}
   \Hbase = \dd \Bbase - \tfrac18 R^2 A \wedge \tF 
      - \tfrac18 R^2 \tA \wedge F.
\end{equation}
Although $h$ clearly vanishes for the $S^3$ solution, we include it here so that our formulae would also describe the extended geometry for more general $S^1$ fibration backgrounds.


\subsection{Non-geometric frames for $S^3$}
\label{sec:anstaz-version}

We now consider a generalised frame $\{\hE_a^L,\hE_{\bar{a}}^R\}$ on the T-fold $X$, where the right frame has the same form as~\eqref{eq:ER} lifted to $X$ and given the identification~\eqref{eq:Tdpsi}
\begin{equation}
\label{eq:ER-tpsi}
\begin{aligned}
   \hE^R_+
      &= \ee^{\ii\phi} \Big[ 
         \left( 2R^{-1}\der_\theta + \tfrac{1}{2}R \dd\theta \right)
         \\ & \qquad \qquad 
         + \ii\cot\theta \left( 2R^{-1}\der_\phi 
             - \tfrac{1}{2}R \dd\phi \right)
          - \ii\csc\theta \,2R^{-1} \big(  \der_\psi + \der_{\tpsi} \big)
          \Big] , \\
   \hE^R_3
      &= 2R^{-1}\der_\phi + \tfrac{1}{2}R\dd \phi , 
\end{aligned}
\end{equation}
whereas for the left frame we introduce a linear phase dependence on $\psi$ and $\tpsi$ 
\begin{equation}
\label{eq:EL-tpsi}
\begin{aligned}
   \hE^L_+ &= \ee^{-\ii(a\psi + b\tpsi)} \Big[ 
         \left( 2R^{-1}\der_\theta - \tfrac{1}{2}R\dd\theta \right)
         \\ & \qquad \qquad 
            + \ii\csc\theta \left( 2R^{-1}\der_\phi - \tfrac{1}{2}R\dd\phi
            \right)
            - \ii\cot\theta\, 2R^{-1}\big( \der_\psi 
               + \der_{\tpsi} \big)  \Big] , \\
   \hE^L_3 &= 2R^{-1} \big( \der_\psi - \der_{\tpsi} \big) .
\end{aligned}
\end{equation}
We then automatically have the correct orthonormality properties in
the $O(3,3)$ metric $\eta$ given in~\eqref{eq:dft-eta}, namely 
\begin{equation}
   \eta(\hE_A,\hE_B)
      = \begin{pmatrix}
            \delta_{ab} & 0 \\
            0 & - \delta_{\bar{a}\bar{b}}
        \end{pmatrix} . 
\end{equation}
Since $\hE^L_{\ba}$ is an $\SO(3)$ rotation of $\hF^L_{\ba}$, (and $\hE^R_a=\hF^R_a$) the generalised metrics $G$ defined by $\{\hE_A\}$ and $\{\hF_A\}$ are the same. In particular, $G$ is independent of $\psi$ and $\tpsi$ and, reducing on $\tpsi$ is simply the round metric on $S^3$ with $H$ flux given in~\eqref{eq:S3pq-soln}.  

It is important to note that this frame explicitly fails to satisfy the weak section condition since generically it depends explicitly on both $\psi$ and $\tpsi$. Thus although the generalised metric is still geometric the particular choice of frame is not. The fact that different global frames on same background correspond to different gaugings is not unusual: for example, viewed as a group manifold, the standard Scherk--Schwarz reduction on $S^3$ defines geometrical frame with an $\SU(2)$ gauge group~\cite{lsw1}, rather than $\SO(4)$. This only implies that the different seven-dimensional gauged supergravity theories admit the same seven-dimensional background as a solution. It is important to note that the scalars of the Scherk--Schwarz reduction define a family of frames of the form $\hE'_A=U_A{}^B\hE_B$, as in~\eqref{eq:SS} but with $U_A{}^B\in\SO(4)$. The corresponding family of generalised metrics $G'$ is given by~\eqref{eq:G'}. For the non-geometrical frames
described here, the generic scalar-dependent generalised metric $G'$ is non-geometric, and violates the weak constraint. It is only at the special point in scalar moduli space where $U_A{}^B=\delta_A{}^B$ that $G$ becomes geometrical. Thus the full uplift of the gauged supergravity is non-geometric, and, because each frame is different, for each value of $a$ and $b$ we get a different uplift. 

We can now calculate $\Lgen_{\hE_A}\hE_B$ using the generalised Lie derivative~\eqref{eq:Lgen-coord}. Since we are violating the section condition there is no reason to believe that this will form an algebra~\cite{dft}. However, we find that
\begin{equation}
\label{eq:ab-cond}
   \Lgen_{\hat{E}^R_a} \hat{E}^L_{\ba} 
      = \Lgen_{\hat{E}^L_{\ba}} \hat{E}^R_a = 0  
      \qquad \text{if} \qquad a + b = 1 , 
\end{equation}
and in addition 
\begin{equation}
\begin{aligned}
\label{eq:S3-ab-alg}
   \Lgen_{\hat{E}^R_a}\hat{E}^R_b 
      &= 2R^{-1} \epsilon_{abc} \hat{E}^R_c \\
   \Lgen_{\hat{E}^L_\ba}\hat{E}^L_\bb 
      &= 2(a - b)R^{-1} \epsilon_{\ba\bb\bc} \hat{E}^L_\bc
\end{aligned}
\end{equation}
and hence we realise the $\SO(4)$ gauge algebra~\eqref{eq:so4} with 
\begin{equation}
   \cot\omega = a - b . 
\end{equation}
Thus the frame gives a non-geometric uplift of the generic gaugings.

We must also consider whether the new frame~\eqref{eq:EL-tpsi} is globally
well-defined. Given the periodicities~\eqref{eq:psi-period}, we see that the factor $\ee^{-\ii(a\psi + b\tpsi)}$ is single valued provided $2a$ and $2b/q$ are integers. For half-maximal supersymmetry, we require not only the frame to be single valued but also the corresponding Killing spinors. Since the frame is constructed as Killing spinor bilinears, the spinors have a phase factor $\ee^{-\ii(a\psi + b\tpsi)/2}$ and hence to be globally defined require
\begin{equation}
   a = m , \qquad  b/q = n ,
\end{equation}
with $m,n \in \bbZ$. The requirement $a+b=1$ gives $m=1-qn$ and hence we get a discrete set of allowed values of $\omega$
\begin{equation}
   \cot \omega = 1 - 2qn \qquad \text{or} \qquad
   \left|\cot\omega\right| = 1, 2q\pm 1, 4q\pm 1, 6q\pm 1, \dots . 
\end{equation}
Note that the algebra with $\omega$ can be mapped to that with $-\omega$ by the $O(6,6)$ transformation that reverses the signs on $\hE^L_\ba$. Hence we can view these solutions as all giving allowed values of $\omega$ lying in the range $0\leq\omega\leq\frac{1}{4}\pi$ with $1\leq\cot\omega\leq\infty$. 

We end by noting first that we can generalise this construction by considering not $S^3$ with $q$ units of flux but the Lens space $S^3/\bbZ_p$ with $q$-units of flux. This is T-dual to the space $S^3/\bbZ_q$ with $p$-units of flux, and corresponds to the periodicities 
\begin{equation}
   \psi \sim \psi + 4\pi/p, \qquad 
   \tpsi \sim \tpsi + 4\pi/q . 
\end{equation}
The conditions that the Killing spinors are globally defined then become
\begin{equation}
   a/p = m , \qquad  b/q = n  
\end{equation}
with $m,n \in \bbZ$ and the allowed values are 
\begin{equation}
   \cot\omega = 1 - 2qn \qquad \text{with} \qquad pm + qn = 1 . 
\end{equation}
which gives a narrower range of allowed values of $\omega$ than in the $S^3$ case. 

Secondly, we note that we can also view the non-geometry from a slightly different perspective. The frame $\{\hE_A\}$ actually only depends on the linear combination $a\psi+b\tpsi$ rather than $\psi$ and $\tpsi$ separately. If we make a change of variables from $(\psi,\tpsi)$ to $(\chi,\tpsi)$ where 
\begin{equation}
\label{eq:sl2z-coord}
   \chi = a\psi + b \tpsi , 
\end{equation}
we see that the $T^2$ fibre of~\eqref{eq:X4} is spanned by the periodicities
\begin{equation}
   \chi \sim \chi + 4\pi, \qquad 
   \tpsi \sim \tpsi + 4\pi/q ,
\end{equation}
and we have simply made an $\SL(2,\bbZ)$ transformation of the original torus. Thus if we quotient by $\der_{\tpsi}$ we see that the coordinates $(\theta,\phi,\chi)$ parametrise a three-sphere, let us call it $\hat{S}^3$. Thus an alternative picture of the geometry is that we consider a global frame on $\hat{S}^3$ with a generalised tangent space
\begin{equation}
\begin{aligned}
   \hat{E} &\simeq T\hat{S}^3\oplus T^*\hat{S}^3 , 
\end{aligned}
\end{equation}
but with a non-standard twisting, $O(3,3)$ metric and bracket, which can be derived by expressing the formulae of section~\ref{sec:T-fold} in terms of the transformed coordinate~\eqref{eq:sl2z-coord}. From this viewpoint, the construction is a generalised parallelisation in a modified generalised geometry on $\hat{S}^3$.


\subsection{Alternative derivation}
\label{sec:s3-gen}

Returning to the original T-fold picture, let us give an alternative derivation of the algebra~\eqref{eq:so4} that will be useful when we turn to the $\SO(8)$ gaugings in the next section. Recall that the structure constants $X_{AB}{}^C$ of the algebra~\eqref{eq:so4} can always be interpreted~\cite{csw2} as the torsion of the generalised Weitzenb\"ock connection $\hat{D}$ defined by the frame $\{\hE_A\}$. Given a global frame, the Weizenb\"ock connection is the unique connection satisfying $\hat{D}\hE_A=0$. By definition
\begin{equation}
\label{eq:XT}
   \Lgen_{\hE_A} \hE_B 
      = \Lgen^{\hat{\Dgen}}_{\hE_A} \hE_B  - T^{\hat{\Dgen}}(\hE_A) \cdot \hE_B
		= - (T^{\hat{\Dgen}}){}_A{}^C{}_B \hE_C , 
\end{equation}
where $(T^{\hat{D}})_A{}^B{}_C$ are the components of the generalised
torsion of the Weizenb\"ock connection evaluated in the $\{\hE_A\}$ frame. Hence 
\begin{equation}
   X_{AB}{}^C = - (T^{\hat{\Dgen}})_A{}^C{}_B . 
\end{equation}
Suppose we have two frames related by an $O(d,d)$ rotation
\begin{equation}
\label{eq:Urot}
   \hE_A = U^B{}_A \hF_B , 
\end{equation}
If $\hat{D}'$ and $\hat{D}$ are the Wietzenb\"ock connections for $\{\hF_A\}$ and $\{\hE_A\}$ respectively then, by defintion,  
\begin{equation}
   \hat{\Dgen} = \hat{\Dgen}' + K  
   \qquad \text{with} \qquad
   K^A{}_B = - (U^{-1}){}^A{}_C \dd U^C{}_B
\end{equation}
and hence the structure constants for the $\{\hE_A\}$ algebra are given by\footnote{Note that here we have defined the index ordering as $T(V){}^A{}_B = V^C T_C{}^A{}_B$, which explains the minus sign in the $K_{[ABC]}$ term with respect to the expression in~\cite{csw1}.} 
\begin{equation}
\label{eq:X'}
\begin{aligned}
   X_{AB}{}^C
      &= - (T^{\hat{\Dgen}})_A{}^C{}_B \\
      &= - (T^{\hat{\Dgen}'})_A{}^C{}_B 
            + 3K_{[ABD]}\eta^{CD} - K_D{}^D{}_A\delta_B{}^C \\
      & = Y'_{AB}{}^C + 3K_{[ABD]}\eta^{CD} - K_D{}^D{}_A\delta_B{}^C , 
\end{aligned}
\end{equation}
where 
\begin{equation}
   Y'_{AB}{}^C = U^D{}_A U^E{}_B (U^{-1})^C{}_F Y_{DE}{}^F, 
\end{equation}
are the structure constants for the $\{\hF_A\}$ algebra transformed to the $\{\hE_A\}$ frame.  

Suppose now that $\{\hF_A\}$ is the standard $\SO(4)$ frame on $S^3$ given by~\eqref{eq:FLR}, that corresponds to $\omega=\frac{1}{4}\pi$, and so, from~\eqref{eq:so4}, gives
\begin{equation}
\label{eq:Y-comp}
   Y_{ab}{}^c = R^{-1}\epsilon_{ab}{}^c , \qquad 
   Y_{\ba\bb}{}^{\bc} = R^{-1}\epsilon_{\ba\bb}{}^{\bc} , 
\end{equation}
with all other components vanishing, and where we have split $A$ into $(a,\ba)$ indices. Now consider a rotated frame of the form~\eqref{eq:Urot}, where 
\begin{equation}
\label{eq:U1}
   U^a{}_b = \delta^a{}_b , \qquad
   U^{\ba}{}_{\bb} 
      = \begin{pmatrix} 
            \cos f & \sin f & 0 \\ 
            -\sin f & \cos f & 0 \\
            0 & 0 & 1
         \end{pmatrix} , 
\end{equation}
and with $U^a{}_\bb=U^\ba{}_b=0$. This is a $U(1)\subset \SO(3)_L\subset \SO(3)_R\times\SO(3)_L$ rotation. As such it does not change the generalised metric $G$. The $\{\hE^R_a\}$ frame is unchanged, as is $\{\hE^L_3\}$. The only frame that transforms is 
\begin{equation}
   \hE^L_+ = \ee^{\ii f} \hF^L_+ . 
\end{equation}
The only non-vanishing component of $K^A{}_B$ is 
\begin{equation}
   K^{\ba}{}_{\bb} = - \dd f \begin{pmatrix} 
                  0 && 1 && 0 \\ -1 && 0 && 0 \\ 0 && 0 && 0 
                  \end{pmatrix}
     := -(\dd f) \lambda^{\ba}{}_{\bb} ,
\end{equation}
where $\lambda^{\ba}{}_{\bb}$ is the generator of the $U(1)\subset\SO(4)$ symmetry. 

To calculate $X_{AB}{}^C$ we need $\dd f\in E^*$ in terms of generalised frame indices. By definition $(\dd f)_A=\hE_A\cdot \dd f=\hE_A(f)$ where, if $v$ is the vector part of $V\in E$, we define $V(f)=v(f)$. On the T-fold $X$ the original frame has the form 
\begin{equation}
\label{eq:F-tpsi}
\begin{aligned}
   \hF^R_+
      &= \ee^{\ii\phi} \Big[ 
         \left( 2R^{-1}\der_\theta + \tfrac{1}{2}R \dd\theta \right)
         \\ & \qquad \qquad 
         + \ii\cot\theta \left( 2R^{-1}\der_\phi 
             - \tfrac{1}{2}R \dd\phi \right)
          - \ii\csc\theta \,2R^{-1} \big(  \der_\psi + \der_{\tpsi} \big)
          \Big] , \\
   \hF^R_3
      &= 2R^{-1}\der_\phi + \tfrac{1}{2}R\dd\phi , \\
   \hF^L_+ &= \ee^{-\ii\psi} \Big[ 
         \left( 2R^{-1}\der_\theta - \tfrac{1}{2}R\dd\theta \right)
         \\ & \qquad \qquad 
            + \ii\csc\theta \left( 2R^{-1}\der_\phi - \tfrac{1}{2}R\dd\phi
            \right)
            - \ii\cot\theta\, 2R^{-1}\big( \der_\psi 
               + \der_{\tpsi} \big)  \Big] \\
   \hF^L_3 &= 2R^{-1} \big( \der_\psi - \der_{\tpsi} \big) . 
\end{aligned}
\end{equation}
We now choose 
\begin{equation}
\label{eq:S3-f-properties}
   \hF_A(f) 
      = \begin{cases} 2c/R & \text{for $\hF^L_3$} \\
         0 & \text{otherwise}
         \end{cases} .
\end{equation}
From~\eqref{eq:F-tpsi}, this implies, up to an irrelevant constant,  
\begin{equation}
   f = \tfrac12 c (\psi - \tpsi) . 
\end{equation}
Given the form of the rotation~\eqref{eq:U1}, we immediately see that we also  have $\hE^L_3(f)=2c/R$ with all other $\hE_A(f)$ vanishing. 

We can now use~\eqref{eq:X'} to calculate the new embedding tensor. We first note that since $U^A{}_B$ is an element of $\SO(3)\times\SO(3)$ we see from~\eqref{eq:Y-comp} that $Y'_{AB}{}^C=Y_{AB}{}^C$. We also note that $K_D{}^D{}_A=0$ and hence $X_{AB}{}^B=0$. This is the analogue of the ``unimodular condition'' for standard Scherk--Schwarz reductions~\cite{lsw1}. We then find
\begin{equation}
   X_{ab}{}^c = 2R^{-1} \epsilon_{ab}{}^c , \qquad
   X_{\ba\bb}{}^\bc = 2R^{-1} (1-c) \epsilon_{\ba\bb}{}^\bc ,
\end{equation}
with all other components vanishing. We see that we reproduce the algebra~\eqref{eq:so4} with $\cot\omega=1-c$. For the Killing spinors corresponding to the new frame $\hE_A$ to be globally well defined, we require the transformation $U$ to be single valued when acting on spinors. This implies $c=2nq$ with $n\in\bbZ$, as before.


\section{$S^7$ and non-geometric $\SO(8)$ gaugings}
\label{sec:s7}


We now address the problem of lifting the general $\SO(8)$ gaugings of~\cite{Dall'Agata:2012bb}. From the proof of section~\ref{sec:strong-violation} we know that any such lifting must violate the section condition. As such the lifts we construct here, though reproducing the correct algebra, will be poorly motivated. In fact, as we will see, even interpreting the extended geometry physically is difficult. Our philosophy is simply to show that there is a self-consistent extended geometry in which a simple section violating construction exists. 


\subsection{The extended geometry}
\label{sec:ext-geom}

Again we will try to consider the minimal possible extension of the geometry. A fully extended space would be 56-dimensional, with a coordinate for each direction in the generalised tangent space~\eqref{eq:E7E}. As discussed in section~\ref{sec:dft}, the requirement that the full extended space admits a global, constant $E_{7(7)}$ structure, along with a suitable foliation, is very restrictive. Here, in direct analogy to the construction in section~\ref{sec:T-fold}, we will only introduce a small number of additional coordinates, in this case four. Viewing $S^7$ as a Hopf fibration over $\CP^3$, we define the ten-dimensional space
\begin{equation}
\label{eq:X10}
   \begin{CD}
      T^4 @>>> X \\
      @. @V{\pi}VV \\
      @. B=\CP^3
   \end{CD}  
\end{equation}
The four angular coordinates $(\psi_0,\psi_2,\psi_4,\psi_6)$ on the $T^4$ fibration are defined by identifying 
\begin{equation}
\label{eq:gen-coords}
\begin{aligned}
   R^{-1} \tfrac{\der}{\der \psi} 
      &= R^{-1} \tfrac{\der}{\der \psi_0} , & && &&
   R^2 \omegaB &= R^{-1} \tfrac{\der}{\der \psi_2} , \\
   \tfrac12 R^5 \dd \psi \wedge \omegaB^2 
      &= R^{-1} \tfrac{\der}{\der \psi_4}, & && &&
   R \dd \psi \otimes \vol_{S^7} &= R^{-1} \tfrac{\der}{\der \psi_6} .
\end{aligned}
\end{equation}
where $\omegaB$ is the standard K\"ahler form on $\CP^3$, normalised such that the volume form on $S^7$ is given by $\vol_{S^7}=\frac{1}{3!}R^7\dd\psi\wedge\omegaB^3$ as discussed in appendix~\ref{app:hopf}. 

If we compactify to type IIA by reducing along $\psi$, we see that $\psi_0=\psi$ is dual to the D0-brane charge on $\CP^3$, while it would appear that $\psi_2$ is related to D2-branes wrapped on $\CP^1\subset \CP^3$, $\psi_4$ to D4-branes wrapped on $\CP^2\subset\CP^3$ and $\psi_6$ to D6-branes wrapping $\CP^3$. In fact as we will discuss in section~\ref{sec:global}, one can use this picture to determine the periodicities of angular coordinates. However, in contrast to the $S^3$ case discussed in section~\ref{sec:T-fold}, this picture is not really consistent. In constrast to the D0-branes, the moduli space of the D2-branes on $\CP^3$ is not itself $\CP^3$, so there is no justification for viewing the wrapping number of the D2-brane charges as corresponding to Kaluza--Klein modes for a $S^1_{\psi_2}$ circle fibered over $\CP^3$. The same is true for the D4- and D6-branes. Similarly, again in constrast to the $S^3$ discussion, there is no duality symmetry exchanging the different brane states.

Notwithstanding these problems of interpretation, one can define a sensible $\E7\times\bbR^+$ extended geometry on $X$. The generalised tangent space has the form
\begin{equation}
   E \simeq TX \oplus \pi^*P^2 \oplus \pi^*P^2 
       \oplus \pi^*T^*B
       \oplus \pi^*\Lambda^5T^*B
       \oplus \pi^*(T^*B\otimes\Lambda^6T^*B) ,
\end{equation}
where $P^n\subset \Lambda^nT^*B$ is the bundle of primitive $n$-forms on the base $B=\CP^3$ (recall an $n$-form $\alpha$ is primitive if $(\omegaB^{-1})^{mn}\alpha_{mnp_1\dots p_{n-2}}=0$).   The components of $V\in E$ can be matched to those in the conventional generalised tangent space on $M=S^7$, as given in~\eqref{eq:E7E}, as follows. Consider the two-form $\omega\in\Lambda^2T^*U_i$, the five-form $\sigma\in\Lambda^5T^*U_i$ and the tensor $\tau\in T^*U_i\otimes \Lambda^7T^*U_i$ on a patch $U_i\subset M$ and expand them as 
\begin{equation}
\begin{aligned}
   \omega &=  \lambda_2 R^2 \omegaB + \beta_1 +  \gamma \wedge \dd\psi, \\
   \sigma &= \big( \tfrac12 \lambda_4 R^5 \omegaB + \beta_2 \big)
      \wedge \omegaB \wedge \dd\psi + \varphi, \\
   \tau &= \lambda_6 R \dd \psi \otimes \vol_{S^7}
      + j \rho \wedge \dd\psi , 
\end{aligned} 
\end{equation}
where $\beta_{1,2}$ are primitive two-forms, and $i_{\der_\psi}\beta_{1,2}=0$. Using~\eqref{eq:gen-coords}, we then identify $\lambda_2$, $\lambda_4$ and $\lambda_6$ with the vector components $v^{\psi_2}$, $v^{\psi_4}$ and $v^{\psi_6}$ of $v\in TX$ respectively, while $\gamma$ is in $\pi^*T^*B$, $\varphi$ is in $\pi^*\Lambda^5T^*B$ and $\rho$ is in $\pi^*(T^*B\otimes\Lambda^6T^*B)$, while $\beta_{1,2}$ are in the two $\pi^*P^2$ parts of $E$. 
As in the $S^3$ case, they imply that $E$ is actually defined as an extension, and that the components $\beta_{1,2}$, $\gamma$ and so on, are not globally forms. Choosing suitable connections, one can always define the corresponding ``untwisted'' objects as in~\eqref{eq:untwist}, which are global forms.

For what follows we need two results. First, the $\E7\times\bbR^+$ structure on $E$ follows directly, using the identifications~\eqref{eq:gen-coords}, from the corresponding structure on the conventional generalised tangent space~\eqref{eq:E7E}. Similarly, the generalised Lie derivative takes the standard form form~\eqref{eq:Lgen-e7}, namely,   
\begin{equation}
\label{eq:Lgen-X}
   \Lgen_V V' = (V\cdot\der)V' - (\der \oadj V)V' , 
\end{equation}
where the projection $\oadj$ uses the aforementioned $\E7\times\bbR^+$ structure and we have  
\begin{equation}
   \der_M = (\der_\alpha,\der_{\psi_0},\der_{\psi_2},
               \der_{\psi_4},\der_{\psi_6},0,\dots,0) .
\end{equation}
where, given coordinates $x^\alpha$ on $\CP^3$, we use $x^m=(x^\alpha,\psi_0,\psi_2,\psi_4,\psi_6)$ coordinates on $X$. Although we will not give the expressions here, as in section~\ref{sec:T-fold} by using the untwisted objects both the $\E7$ invariants and the generalised Lie derivative can be written a manifestly covariant form, dependent on the connections on the four $U(1)$ bundles, which in turn are determined by the flux $\tA$ and the Hopf fibration of $S^7$.

\subsection{Non-geometric frames for $S^7$}
\label{sec:frame}

We construct a new non-geometric frame $\hE_A$ on $S^7$ from the geometrical one~$\hat{F}_A$ defined in~\eqref{eq:E7para}, following the procedure discussed for $S^3$ in section~\ref{sec:s3-gen}. Many of the details are relegated to appendices~\ref{app:s7} and~\ref{app:geom-SU8-indices}. The generalised Weitzenb\"ock connection is used to calculate the corresponding algebra of generalised Lie derivatives using~\eqref{eq:XT}. Concretely we have 
\begin{equation}
\label{eq:S7-rotated-frame}
   \hE_A = U^B{}_A \hF_B , 
\end{equation}
where $U$ is an $\SU(8)$ rotation, and as such does not change the generalised metric, that is the geometry remains a round $S^7$ with $\tF$ flux. The structure constants for $\{\hF_A\}$ and $\{\hE_A\}$ are related by the analogue of~\eqref{eq:X'}, namely, 
\begin{equation}
\label{eq:XY}
   X_{AB}{}^C = Y'_{AB}{}^C + T^{(K)}{}_{AB}{}^C , 
\end{equation}
where again $Y'_{AB}{}^C= U^D{}_A U^E{}_B (U^{-1})^C{}_F Y_{DE}{}^F$, 
\begin{equation}
   K^A{}_B = - (U^{-1}){}^A{}_C \dd U^C{}_B .
\end{equation}
is the difference in the Weitzenb\"ock connections for $\{\hE_A\}$ and  $\{\hF_A\}$, and $T^{(K)}$ is the projection of $K$ onto the torsion representation $\rep{56}+\rep{912}$ of $\E7$.

Decomposing under $\SL(8,\bbR)$, as in~\eqref{eq:X-SL8}, we find that, for the $\SO(8)$ gaugings only the $\rep{36}$ and $\rep{36}'$ representations of $X_{AB}{}^C$ are non-zero
\begin{equation}
   X^{ij} = - R^{-1} \sin\omega \,\delta^{ij} , \qquad 
   \tilde{X}_{ij} = R^{-1} \cos\omega \,\delta_{ij} .
\end{equation}
Equivalently, decomposing under $\SU(8)$ as
\begin{equation}
   \rep{912} = \rep{36} + \rep{\bar{36}} + \rep{420} + \rep{\bar{420}}.
\end{equation}
and comparing with the spinor index form of the algebra~\eqref{eq:alg-SU8}, we find only the $\rep{36}$ and $\bar{\rep{36}}$ parts are non-zero
\begin{equation}
\label{eq:X-spinor}
   X^{\alpha \beta} = R^{-1}\ee^{-\ii\omega} \tC^{\alpha \beta} , \qquad 
   \bar{X}_{\alpha \beta} = R^{-1} \ee^{\ii\omega}\tC_{\alpha \beta} .
\end{equation}
(Note that the vanishing of the $\rep{56}$ component of $X_{AB}{}^C$ is the generalised geometric analogue of the unimodular condition of Scherk--Schwarz reductions~\cite{lsw1}.)

We construct the $U$ rotation as the combination of two separate $\SU(8)$ transformations $U=U''U'$. The first rotation $U'$ exactly follows the construction on $S^3$. Viewing $S^7$ as a Hopf fibration over $\CP^3$ breaks the $\SO(8)$ isometry group to $\SU(4)\times U(1)_H$, where $U(1)_H$ is the action on the Hopf fibre. The first $\SU(8)$ rotation is by this $U(1)_H$ subgroup, and depends on all the extra coordinates $\psi_0$, $\psi_2$, $\psi_4$ and $\psi_6$. The construction is easiest in spinor indices. We have the series of subgroups
\begin{equation}
\begin{array}{ccc}
   \SU(8) & \supset & \Spin(8) \\
   \cup && \cup \\
   \SU(6)\times\SU(2)\times U(1)_C & \ \supset \ & \SU(4)\times U(1)_H
\end{array}
\end{equation}
where $U(1)_H$ is a subgroup of $\SU(2)$. We can see these embeddings explicitly by introducing two real, orthonormal $\Spin(8)$ spinors, invariant under $\SU(6)\subset\SU(8)$ 
\begin{equation}
\label{eq:SU6-basis}
   \eta_1^\alpha
      = \begin{pmatrix} 1 \\ 0 \\ 0 \\ \vdots \\ 0 \end{pmatrix} , 
   \qquad 
   \eta_2^\alpha 
      = \begin{pmatrix} 0 \\ 1 \\ 0 \\ \vdots \\ 0 \end{pmatrix}
   \qquad
   \text{with} \qquad \tilde{C}_{\alpha \beta} = \delta_{\alpha \beta} ,
\end{equation}
such that
\begin{equation}
\label{eq:lambda-def}
   \lambda^\alpha{}_\beta 
      = \eta_1^\alpha \bar\eta_2{}_\beta- \eta_2^\alpha \bar\eta_1{}_\beta
      = \begin{pmatrix} 
         \begin{matrix} 0 & 1 \\ -1 & 0 
            \end{matrix} &  \\  & 0_6 \end{pmatrix} , 
\end{equation}
generates $U(1)_H$, and 
\begin{equation}
   \nu^\alpha{}_\beta 
     = 2\eta_1^\alpha \bar\eta_1{}_\beta + 2\eta_2^\alpha \bar\eta_2{}_\beta
         - \tfrac{1}{2}\delta^\alpha{}_\beta
      = \begin{pmatrix} \tfrac{3}{2}\id_2 & \\ & -\tfrac{1}{2}\id_6 \end{pmatrix}, 
\end{equation}
generates $U(1)_C$. 

Decomposing the frame $\{\hF_A\}= \{\hF_{\alpha\beta},\bar{\hF}^{\alpha\beta}\}$ under $\SU(6)\times\SU(2)\times U(1)_C\subset\SU(8)$ we have 
\begin{equation}
\label{eq:F-decomp}
\begin{aligned}
   \rep{\bar{28}} &= \repp{15}{1}_{-1} + \repp{\bar{6}}{2}_{+1} 
      + \repp{1}{1}_{+3} \\
   \{\hF_{\alpha\beta}\} &= \{ \hF_{\talpha\tbeta} \}
      + \{ \hF_{\talpha 1}, \hF_{\talpha 2} \}
      + \hF_{12}
\end{aligned}
\end{equation}
and similarly for the complex conjugate, where we decompose indices $\alpha=(1,2,\talpha)$. The analogy with the $S^3$ case is that $\hF_{12}$ corresponds to the singlet $\hF^L_3$, while $\hF_{\talpha 1}+\ii\hF_{\talpha 2}$ correspond to $\hF^L_+$, charged under 
 $U(1)_H$, and $\hF_{\talpha\tbeta}$ corresponds to $\hF^R_a$. As there, we consider a $U(1)_H$ rotation, $\hE'_A=U'^B{}_A\hF_B$, expressed in $\SU(8)$ indices as 
\begin{equation}
\label{eq:S7-frame-rotation}
   U^{\prime\alpha}{}_\beta = (\exp f\lambda){}^\alpha{}_\beta 
      = \begin{pmatrix} 
           \cos f & \sin f & \\ -\sin f & \cos f & \\ & & \id_6     
        \end{pmatrix}
\end{equation}
such that $K^\alpha{}_\beta=-(\dd f)\lambda^\alpha{}_\beta$. As in the $S^3$ case we choose the function $f$ such that 
\begin{equation}
\label{eq:S7-f-properties}
   \hE'_{\alpha\beta}(f) = \hF_{\alpha\beta}(f) 
      = \begin{cases} c/R & \text{for $\hE'_{12}=\hF_{12}$} \\
         0 & \text{otherwise}
         \end{cases} .
\end{equation}
Showing that such a suitable $f$ exists is a somewhat lengthy calculation, the details of which are presented in the appendices~\ref{app:s7} and~\ref{app:geom-SU8-indices}. We find 
\begin{equation}
\label{eq:s7-f-def}
   f = a(\psi_0 + 3 \psi_4) + b(\psi_2 + \tfrac13 \psi_6) ,
\end{equation}
giving $c=-\tfrac{1}{12}b - \tfrac14 \ii a$.

We can then calculate the corresponding algebra for $\hE'_A=U^{\prime B}{}_A\hF_B$ using~\eqref{eq:XY}. We first note that since $U'\in\Spin(8)$ we have $Y'_{AB}{}^C=Y_{AB}{}^C$. Taking care to get the normalisations correct we then find that the $\rep{36}$ and $\bar{\rep{36}}$ parts of $T^{(K)}$ are given by (see appendix~\ref{app:geom-SU8-indices})
\begin{equation}
   T^{(K)\alpha \beta} = 2R^{-1}\left( a + \tfrac{1}{3}\ii b\right) 
      L^{\alpha \beta} ,
   \qquad
   \bar{T}^{(K)}{}_{\alpha \beta} = 2R^{-1}\left( a - \tfrac{1}{3}\ii b\right)    
     \bar{L}_{\alpha \beta} ,
\end{equation}
where
\begin{equation}
   L^{\alpha\beta} = 
      \eta_1^\alpha \eta_1^\beta + \eta_2^\alpha \eta_2^\beta
		= \begin{pmatrix} \id_2 &  \\  & 0_6 \end{pmatrix} , 
\end{equation}
and in addition the $\rep{28}$ and $\rep{420}$ (and their complex conjugate) components vanish. If we now fix the constants so that 
\begin{equation}
\label{eq:angle-cond}
   (1- 2a) - \tfrac23 \ii b = \ee^{-4\ii \omega}
\end{equation}
we then have 
\begin{equation}
\label{eq:nongeom-torsion}
\begin{aligned}
   X^{\prime\alpha \beta} 
     &= R^{-1} \left[ \tC^{\alpha \beta} 
         + \left(\ee^{-4\ii \omega}-1\right) L^{\alpha \beta} \right]
     = \frac{1}{R} \begin{pmatrix} 
          \ee^{-4 \ii \omega} \id_2 & \\ & \id_6 \end{pmatrix} , \\
   \bar{X}'_{\alpha \beta} &= R^{-1} \left[ \tC_{\alpha \beta}
	 + \left(\ee^{4\ii \omega}-1\right) \bar{L}_{\alpha \beta} \right]
     = \frac{1}{R} \begin{pmatrix} 
        \ee^{4 \ii \omega} \id_2 & \\ & \id_6 \end{pmatrix} . 
\end{aligned}
\end{equation}

This new embedding tensor $X'_{AB}{}^C$ is actually equivalent to the Dall'Agata et al. tensor~\eqref{eq:X-spinor} under a constant $U(1)_C\subset\SU(8)$ rotation. Explicitly, we make a second transformation $\hE_A=U^{\prime\prime B}{}_A\hE'_B=(U''U')^B{}_A\hF_A$ with 
\begin{equation}
\label{eq:S7-frame-rotation-2}
   U^{\prime\prime\alpha}{}_\beta = (\exp [-\ii\omega\nu]){}^\alpha{}_\beta 
      = \begin{pmatrix} 
           \ee^{-3\ii\omega/2}\id_2 & \\ 
           & \ee^{\ii\omega/2} \id_6     
        \end{pmatrix} .
\end{equation}
Since the transformation is constant the corresponding $K^A{}_B$ vanishes and from~\eqref{eq:XY} we have the new embedding tensor
\begin{equation}
   X^{\alpha \beta} = R^{-1}\ee^{-\ii\omega} \tC^{\alpha \beta} , \qquad
   \bar{X}_{\alpha \beta} = R^{-1} \ee^{\ii\omega}\tC_{\alpha \beta} , 
\end{equation}
as for Dall'Agata et al, as required. Using the decomposition~\eqref{eq:F-decomp}, the combined  transformation can be written explicitly as 
\begin{equation}
\label{eq:E-ng}
\begin{aligned}
   \hE_{\talpha \tbeta} 
      &= \ee^{+\ii \omega}  \hF_{\talpha \tbeta} , \\
   \hE_{\talpha\pm} &= \ee^{\ii (-\omega \pm f)}  \hF_{\talpha \pm} , \\
   \hE_{12} &= \ee^{-3\ii \omega}  \hF_{12} ,
\end{aligned}
\end{equation}
where $\hE_{\talpha\pm}=\hE_{\talpha1}\pm\ii \hE_{\talpha2}$. We have found an explicit non-geometrical frame realising the general $\SO(8)$ gauging.


\subsection{Quantisation and global structure}
\label{sec:global}

In section~\ref{sec:ext-geom} we saw that naively the extended space coordinates $\psi_0$, $\psi_2$, $\psi_4$ and $\psi_6$ are associated to D0-, D2-, D4- and D6-branes on $\CP^3$ respectively. Such a correspondence allows us to use the brane charges to fix the periodicities of the corresponding circles, just as in the $S^3$ case, where the winding string states fixed the periodicity of the dual $\tpsi$ circle to be $4\pi/q$. As we discussed, this picture is not really consistent because the moduli spaces of the D2-, D4- and D6-branes are not themselves $\CP^3$. Nonetheless it is the only tool we have to try and fix the relevant periodicities, and it is interesting to see what it implies. 

Following~\cite{Aharony:2009fc,Gutierrez:2010bb}, the masses of the D2$n$-branes wrapping $\CP^n$ cycles in $\CP^3$ are calculated in appendix~\ref{app:charges}, and then converted into radii in M theory units. To keep things as general as possible, we actually include the possibility that the seven-dimensional space is the ABJM~\cite{ABJM} Hopf-fibre quotient $S^7/\bbZ_p$. Using the relations~\eqref{eq:gen-coords}, we match these masses onto Kaluza--Klein masses for the four circular coordinates, and arrive at the periodicities
\begin{equation}
\begin{aligned}
   \psi_0 &\sim \psi_0 + \frac{2\pi}{p} , && &
   \psi_2 &\sim \psi_2 + \beta \frac{2\pi}{p} ,  \\
   \psi_4 &\sim \psi_4 + \frac{2\pi}{q} , && &
   \psi_6 &\sim \psi_6 + 3\beta \frac{2\pi}{q} , 
\end{aligned}
\end{equation}
where $\beta = \sqrt{p/2q}$.

We now see what these periodicities imply for the non-geometric frame~\eqref{eq:E-ng}. From appendix~\ref{app:hopf}, we see that the $\hE_{\talpha\tbeta}$ and $\hE_{12}$ components are independent of all the $\psi_{2n}$ coordinates, but $\hE_{\talpha\pm}$ has the dependence 
\begin{equation}
   \hE_{\talpha\pm} = \ee^{\pm\ii(f+2\psi_0)} \big[ \dots \big] . 
\end{equation}
As in the $S^3$ case the frame is constructed as a Killing spinor bilinear, so for the Killing spinors to be globally defined we need phase factor $\ee^{\ii(f/2-\psi_0)}$ to be single valued. This implies 
\begin{equation}
\label{eq:quant-cond}
   \tfrac{1}{2}a = p n_0 - 1 = \tfrac{1}{3} q n_4 , \qquad
   \tfrac{1}{2}\beta b = p n_2 = q n_6 
\end{equation}
for charges $n_i \in \bbZ$. It is encouraging that we find the second of these relations here, as it is noted in~\cite{ABJM} that the only allowable configurations of $n_2$ $D2$-branes and $n_6$ $D6$-branes in this geometry must satisfy this relation.

The conditions~\eqref{eq:quant-cond} ensure that the frame is globally well-defined on the extended space. However, in order that the frame realises the Dall'Agata et al. algebra, we also had to choose the constants $a$ and $b$ to satisfy~\eqref{eq:angle-cond}. This implies a quadratic relation between the D-brane charges $n_i$ and the geometric data $p$ and $q$, namely,
\begin{equation}
   \left(1 - \frac{4q n_4}{3} \right)^2 
      + 2pq \left( \frac{4n_2}{3} \right)^2 = 1
\end{equation}
It is simple to check that the only solutions for $p,q,n_i \in \bbZ$, have $a=b=0$, which implies $\omega=0$. This conclusion also holds if we require only the frame $\hE_A$ to be globally defined and not the Killing spinors themselves. The non-geometric frames fail to satisfy the desired quantisation conditions. Therefore, we are led to the conclusion that when one uses the only tool we have to fix the global properties of the extended space, there are in fact no new solutions.


\section{Conclusions}
\label{sec:con}

The first part of this paper proved a no-go theorem that the generic $\SO(8)$ gaugings of~\cite{Dall'Agata:2012bb} cannot be realised as a consistent truncation of either eleven-dimensional or type II supergravity, or as truncations of extended spacetimes that satisfy the section condition. This strongly suggests that if they are to be realised in string or M-theory the background must be intrinsically stringy, and so not captured by supergravity, or by T-fold or U-fold backgrounds that still admit a local supergravity description. We also proved a similar no-go theorem for the generic $\SO(4)$ gaugings of half-maximal seven-dimensional supergravity of~\cite{DLR,DFMR}.   

The second part of the paper is much more speculative. Introducing the minimal number of additional coordinates possible, we showed, building on the results of~\cite{lsw1}, that the standard geometric round $S^7$ solution admits a family of non-geometric generalised frames. The frames depend on the additional coordinates in a way that violates the section condition. Assuming the generalised Lie derivative nonetheless takes the standard form, we showed these new frames reproduce the algebra of the generic $\SO(8)$ gaugings. We also showed that an analogous set of generalised frames on $S^3$ reproduces the generic $\SO(4)$ gaugings. In both cases, although at one point the frame defines a conventional $S^d$ geometry, for generic values of the scalar fields in the truncation, the uplifted background is non-geometric. 

By associating the additional coordinates to the charges of corresponding wrapped branes, we where able to fix their periodicities. For the $\SO(4)$ case, requiring the frame to be globally defined restricted the allowed gaugings to a discrete set given by $|\cot\omega|=1,2qn\pm1,4q\pm1,\dots$ where $q\in\bbZ$ labels the number of units of $H$-flux. For the $S^7$ case, we found that there were no global solutions other than that giving the standard $\omega=0$ geometrical $\SO(8)$ gauging of~\cite{deWit:1982ig}. It was shown in~\cite{Dall'Agata:2014ita} that the gauging parameters should be continuous, thus even for the $\SO(4)$ case we seem to be missing some of the possible gaugings. 

These results are consistent with the detailed AdS/CFT analysis  of~\cite{Borghese:2014}. There it was shown that the standard boundary conditions for fluctuations in $AdS_4$ of the generic $\SO(8)$ gaugings are only supersymmetric for the standard $\omega=0$ case. However, there is an $\mathcal{N}=6$ truncation which \emph{classically} is independent of $\omega$, does preserve supersymmetry and simply corresponds to the ABJM background~\cite{ABJM}. From~\eqref{eq:E-ng}, we see that the $\hE_{\talpha\tbeta}$ components of the frame are globally defined and geometrical, independent of all the angular directions $\psi_i$ (in fact they are defined on the quotient $S^7/\bbZ_p$). They are invariant under $\SU(2)\subset \SU(8)$ and hence define a background preserving $\mathcal{N}=6$ supersymmetry. Furthermore, under the generalised Lie derivative they define an $\SO(6)$ gauging. They thus give the uplift the $\mathcal{N}=6$ truncation discussed in~\cite{Borghese:2014}.

As we have stressed, finding a section-violating frame that reproduces the correct gauge algebra is not the same as showing the model can be uplifted to string or M-theory, precisely because an understanding of DFT and its M-theory cousin away from toroidal backgrounds, and furthermore when the section condition is violated, remains an open question. First one needs to define an appropriate extended geometry. As discussed in section~\ref{sec:dft}, this always requires some additional structure on the underlying geometrical space. For the case of $S^3$, it requires identifying a vector field and one-form on $S^3$ corresponding to the Hopf fibration. One can then define a four-dimensional T-fold space by also including the T-dual of the Hopf fibre, and a generalised tangent space which admits a covariant global $O(3,3)$ metric and generalised Lie derivative. The $S^7$ case is under less control. Reducing to type IIA on $\CP^3$ along the Hopf fibre, we introduced addition circle  coordinates that naively were dual to wrapped D2-, D4- and D6-branes along with the D0-branes from the Hopf fibre. However in this case there is no duality transformation relating these fibres. Furthermore, the D2-branes and D6-branes moduli spaces are not $\CP^3$ and so it is not clear how they define a fibration. Furthermore, for both the $S^3$ and $S^7$ cases, the generalised frame violates the weak form of the section condition. In a string context this implies that modular invariance is violated and so is a radical departure from standard string constructions. For these reasons, despite showing that there is a self-consistent way to realise the generic $\SO(8)$ gauge algebra in an extended geometry, we remain sceptical as to whether this provides a way of realising the theory in M-theory. 
This, together with the fundamental problem of defining the additional coordinates for non-toroidal backgrounds, suggests to us that a different approach is needed if the generic gaugings are to be realised in M theory.


\acknowledgments

We would like to thank Mariana Gra\~{n}a, Chris Hull, Michela Petrini and Arkady Tseytlin for helpful discussions. We especially thank Chris Hull for invaluable discussions on the material in section~\ref{sec:dft}. C.~S-C.~has been supported by the German Science Foundation (DFG) under the Collaborative Research Center (SFB) 676 ``Particles, Strings and the Early Universe'' and by a grant from the Foundational Questions Institute (FQXi) Fund, a donor advised fund of the Silicon Valley Community Foundation on the basis of proposal FQXi-RFP3-1321 (this grant was administered by Theiss Research).  D.~W.~is supported by the STFC grant ST/J000353/1 and the EPSRC Programme Grant EP/K034456/1 ``New Geometric Structures from String Theory''. C.~S-C.~would like to thank Imperial College London for hospitality during the completion of this work. D.~W.~also thanks the Berkeley Center for Theoretical Physics for kind hospitality during the final stages of this work.


\appendix


\section{A no-go theorem for $\SO(4)$ gaugings}
\label{app:nogo}


In this appendix we derive a no-go theorem for local geometrical lifts of the $\SO(4)$ gaugings~\eqref{eq:SO4-nongeom-X}. Such a lift requires a frame  $\{\hat{E}_A\}=\{\hat{E}^R_a,\hat{E}^L_{\bar{a}}\}$ satisfying the algebra~\eqref{eq:so4}. The proof of the theorem follows closely that for the $\SO(8)$ given in section~\ref{sec:nogo}.

We can expand the frame 
\begin{equation}
   \hE^L_a = u_a + \dots , \qquad 
   \hE ^R_{\bar{a}} = v_{\bar{a}} + \dots ,
\end{equation}
where the $+\dots$ represent the one-form parts. Since both $\hE^L_a$ and $\hE^R_{\bar{a}}$ form $\su(2)$ algebras and $\su(2)$ is simple, we can again conclude that none of the $u_a$ and $v_{\bar{a}}$ can vanish. Furthermore, given the frame is orthonormal, that is the generalised metric satisfies
\begin{equation}
   G(\hat{E}_A,\hat{E}_B)
      = \begin{pmatrix}
            \delta_{ab} & 0 \\
            0 & \delta_{\bar{a}\bar{b}}
        \end{pmatrix} ,
\end{equation}
we again have that $\{\hat{E}^R_a,\hat{E}^L_{\bar{a}}\}$ are generalised Killing vectors. Crucially the condition $\Lgen_{\hat{E}^L_{\bar{a}}}\hat{E}^R_a=0$ implies that $\BLie{u_a}{v_{\bar{a}}}=0$. Thus we have six non-zero Killing vectors generating the $\so(4)=\su(2)\oplus\su(2)$ algebra. We are forced to conclude that the local geometry is the round $S^3$ with constant $H$-flux, and, up to an irrelevant global $\SO(4)$ rotation we have 
\begin{equation}
   \hE^R_{a} = r_a+ \dots , \qquad 
   \hE^L_{\ba} = \tan\omega\, l_{\ba} + \dots ,
\end{equation}
where $r_a$ and $l_{\ba}$ are the right- and left-invariant vector fields on $S^3$ given in~\eqref{eq:inv-vec}. 

Recall that we showed in~\cite{lsw1} that the round $S^3$ geometry admitted a (global) generalised frame, given in~\eqref{eq:FLR}, such that 
\begin{equation}
   \hF^R_{a} = r_a + \dots , \qquad
   \hF^L_{\ba} = l_\ba+ \dots . 
\end{equation}
Since any other orthonormal frame must be related by a local $\SO(3)\times\SO(3)$ rotation, we need a local $\SO(3)\times\SO(3)$ transformation $U$ such that 
\begin{equation}
   U_{a}{}^{b} v_b = v_{b} , \qquad 
   U_{\ba}{}^{\bb} v_{\bb} =  \tan \omega \, v_{\ba} . 
\end{equation}
Taking the norms of each side in the local $S^3$ metric, it is clear that there is no such $U$ unless $\omega = \frac{1}{4}\pi$, which is the case of the original $S^3$ gauging. As for the new $\SO(8)$ gaugings, we have now shown that there is no way to realise this family of gauge algebras with a local generalised frame.


\section{$S^3$ conventions}
\label{app:s3}

We take the metric on $S^3$
\begin{equation}
   \dd s^2 = \tfrac{1}{4}R^2\left[ \dd\theta^2 
         + \sin^2\theta \dd \phi^2
         + \left(\dd\psi + \cos\theta\dd\phi\right)^2 \right] , 
\end{equation}
where $\psi\simeq\psi+4\pi$. The standard left- and right-invariant vector fields are 
\begin{equation}
\label{eq:inv-vec}
\begin{aligned}
   l_+ &= l_1+\ii l_2 
      = 2R^{-1} \ee^{-\ii\psi} \big[  \der_\theta 
         + \ii\csc\theta \der_\phi - \ii\cot\theta \der_\psi  \big] , 
         & &&
   l_3 &= 2R^{-1} \der_\psi , \\
   r_+ &= r_1 + \ii r_2 
      = 2R^{-1}\ee^{\ii\phi} \big[  \der_\theta 
         + \ii\cot\theta \der_\phi  - \ii\csc\theta \der_\psi \big] , 
         & && 
   r_3  &= 2R^{-1}\der_\phi , 
\end{aligned}
\end{equation}
with the corresponding left- and right-invariant one-forms
\begin{equation}
\label{eq:inv-form}
\begin{aligned}
   \lambda_+ 
     &= \tfrac{1}{2}R \ee^{-\ii\psi} \left(\dd\theta 
         + \ii\sin\theta\dd\phi\right) , 
     & && 
   \lambda_3 &= \tfrac{1}{2}R \left( \dd\psi 
         + \cos\theta\dd\phi \right) , \\
   \rho_+ &= \tfrac{1}{2}R \ee^{\ii\phi} \left(\dd\theta 
         - \ii\sin\theta\dd\psi \right) , 
     & && 
   \rho_3 &= \tfrac{1}{2}R \left( \dd\phi + \cos\theta\dd\psi \right) . 
\end{aligned}
\end{equation}
%


\section{$S^7$ conventions and calculations}
\label{app:s7}


\subsection{Definitions for $\{\hF_A\}$ frame}
\label{app:hF}

We briefly review the definitions of some objects from~\cite{lsw1}
used to construct the generalised frames $\{\hF_A\}$ on $S^7$. We
consider constrained coordinates $y^i$, for $i = 1, \dots, 8$ with
$\delta_{ij} y^i y^j = 1$. In terms of these variables, the round
metric on $S^7$ takes the form
\begin{equation}
\label{eq:S7metric}
   \dd s^2 = R^2 \dd s_{S^7}^2 = R^2 \delta_{ij} \dd y^i \dd y^j .
\end{equation}
The Killing vectors $v_{ij}$ were expressed in terms of the conformal Killing vectors $k_i$ (satisfying $\mathcal{L}_{k^i} g = - 2 y^i g$) by
\begin{equation}
\label{eq:Killing-vecs}
   v_{ij} = R^{-1}\left( y_ik_j - y_jk_i \right) ,
\end{equation}
We also define
\begin{equation}
\label{eq:ost-def}
\begin{aligned}
   \omega_{ij} &= R^2 \dd y_i \wedge \dd y_j , \\
   \sigma_{ij} = {}* \omega_{ij}
      &= \frac{R^5}{5!}\epsilon_{ijk_1\dots k_6}
          y^{k_1}\dd y^{k_2}\wedge \dd y^{k_6} 
          = - i_{k_i} i_{k_j} \vol_{g}, \\
   \tau_{ij} &= R (y_i\dd y_j - y_j\dd y_i)\otimes \vol_g
\end{aligned}
\end{equation}
where
\begin{equation}
\label{eq:vol}
   \vol_g = R^7 \vol_{S^7}= \frac{R^7}{7!}\epsilon_{i_1\dots i_8}
          y^{i_1}\dd y^{i_2}\wedge \dots \wedge \dd y^{i_8} .
\end{equation}
%


\subsection{The Hopf fibration}
\label{app:hopf}

To define the $S^7$ as the Hopf fibration over $\CP^3$ we introduce the complex coordinates\footnote{The minus sign in the definition of $z^4$ here ensures that the orientation of the complex structure matches the $\SU(4)$ decomposition of the spinor representations in appendix~\ref{app:geom-SU8-indices}, see e.g. equation~\eqref{eq:eta-spinor-def}, as we use the negative chirality representation~\eqref{eq:-ve-chirality}.} on $\bbC^4$
\begin{equation}
\begin{aligned}
   z^1 &= y^1+\ii y^2 , & 
   z^2 &= y^3+\ii y^4 , &
   z^3 &= y^5+\ii y^6 , &
   z^4 &= y^7-\ii y^8 . 
\end{aligned}
\end{equation}
The unit $S^7$ sphere is defined by the constraint $z^a\bar{z}_a=1$, where $\bar{z}_a=\delta_{a\bar{a}}\bar{z}^{\bar{a}}$ and $\CP^3$ is the quotient of $S^7$ by the $U(1)$ action $z^a \to \ee^{i \alpha} z^a$. The metric~\eqref{eq:S7metric} can be written as 
\begin{equation}
   \dd s_{S^7}^2 = \dd z^a \dd \bar{z}_a 
      = \lambda^2 + \dd s_{\CP^3}^2 , 
\end{equation}
where 
\begin{equation}
   \lambda = \tfrac{1}{2}\ii \left( 
      z^a \dd\bar{z}_a - \bar{z}_a \dd z^a \right) ,
\end{equation}
and $\dd s_{\CP^3}^2$ is the Fubini--Study metric on $\CP^3$, manifestly invariant under an $SU(4)$ action on $z^a$, 
\begin{equation}
\label{eq:CP3-metric}
   \dd s^2_{\CP^3} = 
      \dz^a \dd\bar{z}_a - (z^a\dd\bar{z}_a)(\bar{z}_b\dd z^b) . 
\end{equation}
Note that one can also construct the corresponding K\"ahler form on $\CP^3$
\begin{equation}
\label{eq:omegaB-def}
   \omegaB = \tfrac{1}{2}\ii \dz^a \wedge \dd\bar{z}_a , 
\end{equation}
such that 
\begin{equation}
\label{eq:vols}
   \vol_{S^7} = \lambda \wedge \vol_{\CP^3} , \qquad
   \vol_{\CP^3} = \tfrac{1}{3!}\omegaB^3  , 
\end{equation}
where throughout we are using $z^a\dd\bar{z}_a+\bar{z}_a\dd z^a=0$ as a result of the constraint $z^a\bar{z}_a=1$. 

Introducing the unconstrained coordinates
\begin{equation}
\label{eq:CP3-coord}
\begin{aligned}
   z^i &= r \ee^{\ii\psi} w^i, & i &= 1,2,3 & && && &&
   z^4 = r \ee^{\ii\psi} ,
\end{aligned}
\end{equation}
where $r^2 (1 + w^i \bw_i) = 1$ on $S^7$, one has the standard expressions
\begin{equation}
\label{eq:A-def}
   \lambda = \dd \psi + A  , \qquad
   A = \tfrac{1}{2}\ii r^2 (w^i \dwb_i - \bw_i \dw^i) ,
\end{equation}
and 
\begin{equation}
   \omegaB = \tfrac{1}{2}\ii \left( 
      r^2 \delta_i{}^j - r^4 \bw_i w^j \right) \dw^i \wedge \dwb_j .
\end{equation}
We can then see explicitly that $\dd A = 2 \omegaB$.

In terms of the real coordiantes $y^i$, we note that 
\begin{equation}
\begin{aligned}
   \lambda &= \tfrac{1}{2}\Omega_{ij} y^i\dd y^j , & && &&
   \omegaB &= \tfrac{1}{2}\Omega_{ij} \dd  y^i \wedge \dd y^j , 
\end{aligned}
\end{equation}
where $\Omega$ is the $\SU(4)$-invariant symplectic form on $\bbC^4$, with the non-zero components $\Omega_{12}=\Omega_{34}=\Omega_{56}=-\Omega_{78}=1$. Hence for the objects defined in appendix~\ref{app:hF}, one finds 
\begin{equation}
\label{eq:singlets}
\begin{aligned}
   \tfrac{1}{2}\Omega_{ij} v^{ij} &= R^{-1}\der_\psi , & && &&
   \tfrac{1}{2}\Omega_{ij} \omega^{ij} &= R^2\omegaB , \\
   \tfrac{1}{2}\Omega_{ij} \sigma^{ij} &= * \omegaB 
      = R^5\lambda\wedge \tfrac{1}{2}\omegaB^2 , & && && 
   \tfrac{1}{2}\Omega_{ij} \tau^{ij} &= R\lambda\otimes \vol_g .
\end{aligned}
\end{equation}
%


\section{$E_{7(7)} \times \bbR^+$ generalised geometry in $\SU(8)$ indices}
\label{app:geom-SU8-indices}

In this appendix we provide expressions for $E_{7(7)} \times \bbR^+$ generalised geometry objects expressed in $\SU(8)$ indices, following the conventions of~\cite{csw2,csw3}. 


\subsection{Gamma matrix conventions and formulae}
\label{app:gammas}

We follow the conventions of~\cite{csw3}. Briefly, these include taking the representation of $\Cliff(7,\bbR)$ with $\gamma^{(7)} = -\ii$ and building a representation of the $\Spin(8)$ algebra via
\begin{equation}
\label{eq:gamma8}
   \gamh^{ij} = \begin{cases} 
          \gamma^{ab} & i=a , j=b \\
          +\gamma^{a} \gamma^{(7)} & i=a, j=8 \\
          -\gamma^b \gamma^{(7)} & i = 8, j=b 
       \end{cases} ,
\end{equation}
This representation has negative chirality as
\begin{equation}
\label{eq:-ve-chirality}
   \gamh^{i_1 \dots i_8} = - \epsilon^{i_1 \dots i_8} .
\end{equation}
We have the useful completeness relations,
\begin{equation}
\label{eq:gamma-completeness}
\begin{aligned}
   \gamh^{ij}{}_{\alpha \beta} \gamh_{ij}{}^{\gamma \delta} 
      = 16 \delta^{\gamma \delta}_{\alpha \beta} ,
      \hs{50pt}
   \gamh^{ij}{}_{\alpha \beta} \gamh_{kl}{}^{\alpha \beta} 
      = 16 \delta^{ij}_{kl} ,
\end{aligned}
\end{equation}
where $\delta_{ij}^{kl}=\delta_{[i}^k\delta_{j]}^l$ and we use the transpose intertwiner $\tilde{C} = \tilde{C}^T$ to raise and lower spinor indices, and a Fierz identity, which also serves as our definition of $\epsilon_{\alpha_1 \dots \alpha_8}$, 
\begin{equation}
   \tfrac{1}{4!} 
     \epsilon_{\alpha \alpha' \beta \beta' \gamma \gamma' \delta \delta'}
     \gamh^{ij \gamma \gamma'} \gamh^{kl \delta \delta'}
     = 2 \gamh^{[ij}{}_{[\alpha \alpha'} \gamh^{kl]}{}_{\beta \beta']}
        - \gamh^{ij}{}_{[\alpha \alpha'} \gamh^{kl}{}_{\beta \beta']} .
\end{equation}
Another Fierz identity we will need is
\begin{equation}
	\gamh^{[i}{}_k{}_{\alpha\alpha'} \gamh^{j]k}{}_{\beta\beta'}
		= - 4 \tilde{C}_{\alpha \beta} \gamh^{ij}{}_{\beta' \alpha'}
\end{equation}
where we use the antisymmetrisation convention that $\alpha \alpha'  \equiv [\alpha \alpha']$.


\subsection{Index conventions}
\label{app:index}

First we split the $E_{7(7)}\times\bbR^+$ frame index $A = 1, \dots, 56$ into a pair of antisymmetrised indices under the $\SL(8,\bbR)$ subgroup as 
\begin{equation}
   V = V^A \hE_A = \tfrac{1}{2}\big( 
      V^{ii'} \hE_{ii'} + \tV_{ii'} \hE'^{ii'} \big) \in E.  
\end{equation}
An example of such a frame is the ``conformal split frame''
\begin{equation}
\label{eq:csf}
\begin{aligned}
   \hE_{a8} &= \hE_a , & && && 
   \hE_{ab} &= \tfrac{1}{5!} \epsilon_{abc_1\dots c_5} \hE^{c_1\dots c_5} , \\
   \hE^{\prime a8} &= \tfrac{1}{7!} 
      \epsilon_{b_1\dots b_7} \hE^{a,b_1\dots b_7} , & && && 
   \hE^{\prime ab} &= \hE^{ab} ,
\end{aligned}
\end{equation}
where we split $i=(a,8)$ and define
\begin{equation}
\label{eq:geom-basis}
\begin{aligned}
   \hE_a &= \ee^{\Delta} \Big( \hat{e}_a + i_{\hat{e}_a} A
      + i_{\hat{e}_a}\tA 
      + \tfrac{1}{2}A\wedge i_{\hat{e}_a}A 
      \\ & \qquad \qquad 
      + jA\wedge i_{\hat{e}_a}\tA 
      + \tfrac{1}{6}jA\wedge A \wedge i_{\hat{e}_a}A \Big) , \\
   \hE^{ab} &= \ee^\Delta \left( e^{ab} + A\wedge e^{ab} 
      - j\tA\wedge e^{ab}
      + \tfrac{1}{2}jA\wedge A \wedge e^{ab} \right) , \\
   \hE^{a_1\dots a_5} &= \ee^{\Delta} \left( e^{a_1\dots a_5} 
      + jA\wedge e^{a_1\dots a_5} \right) , \\
   \hE^{a,a_1\dots a_7} &= \ee^\Delta e^{a,a_1\dots a_7} . 
\end{aligned}
\end{equation}
given a conventional frame $\{\hat{e}_a\}$ for $TM$ and its dual $\{e^a\}$ for $T^*M$. Conformal split frames always exist, and any other frame is related to~\eqref{eq:csf} by a general local $\E7\times\bbR^+$ transformation. 

Given two generalised vectors $V,V'\in E$ of the form~\eqref{eq:E7E}, the $\E7$ symplectic invariant defines a top-form 
\begin{equation}
   \bl{V}{V'} = 
      \tfrac{1}{2}\left(i_v\tau' -i_{v'}\tau 
         + \sigma\wedge\omega' - \sigma'\wedge\omega \right) 
      \in \Lambda^7T^*M ,
\end{equation}
where $(i_v\tau)_{m_1\dots m_7}=v^m\tau_{m,m_1\dots m_7}$. The $\E7\times\bbR^+$ frame $\{\hE_A\}$ has
\begin{equation}
\label{eq:pairSL8}
   \bl{\hE_{ij}}{\hE^{\prime kl}}
      = \Phi^2\delta_{ij}^{kl} , \qquad
   \bl{\hE_{ij}}{\hE_{kl}} = 0 , \qquad
   \bl{\hE^{\prime ij}}{\hE^{\prime kl}} = 0 , 
\end{equation}
where $\Phi^2\in\Lambda^7T^*M$ is a volume form that depends on the choice of frame, and scales under the action of $\bbR^+$. For a split conformal frame it is given by $\Phi^2=\ee^{2\Delta}e^{1\dots 7}$. In terms of the generalised metric $G$, defined by $\{\hE_A\}$ via~\eqref{eq:GSL8}, it is $\vol_G=\ee^{2\Delta}\vol_g$. 

The symplectic invariant (together with the density $\Phi$) defines an isomorphism between $E$ and $E^*$, and so we can introduce a similar decomposition for the dual basis $\{E^A\}\in E^*$
\begin{equation}
   W = W_A E^A = \tfrac{1}{2}\big( 
      W^{ii'} E_{ii'} + \tilde{W}_{ii'} E'^{ii'} \big) \in E^* ,   
\end{equation}
where $E_{ii'}=-2\Phi^{-2}\hE_{ii'}$ and $E'^{ii'}=2\Phi^{-2}\hE'^{ii'}$, so that, for example, for the conformal split frame, we have $E'^{a8}=2\ee^{-\Delta}e^a$. The factor of two is conventional, and implies the contraction between $V\in E$ and $W\in E^*$ is given by
\begin{equation}
\label{eq:contract}
   V\cdot W = V^A W_A = V^{ii'} \tilde{W}_{ii'} + \tV_{ii'}W^{ii'} . 
\end{equation}
In particular the derivative along $V$ is given by 
\begin{equation}
   \der_V = V^A\der_A = V^{ii'} \der_{ii'} + \tV_{ii'} \tder^{ii'} = v^m\der_m ,   
\end{equation}
where, in the conformal split frame, if $V$ has the form~\eqref{eq:E7E} then  $V^{a8}=\ee^{-\Delta}v^a$ and $\der_{a8}=\frac{1}{2}\ee^\Delta\der_a$. 

One can similarly decompose $\hE_A$ under $\SU(8)$ as 
\begin{equation}
\label{eq:V-su8-basis}
   V = V^A \hE_A = \tfrac12\big(
        V^{\alpha\alpha'} \hE_{\alpha\alpha'} 
        + \bV_{\alpha\alpha'} \bhE^{\alpha\alpha'} \big) ,
\end{equation}
where $\bhE^{\alpha\alpha'}$ is the complex conjugate of $\hE_{\alpha\alpha'}$. We use the common $\SO(8)$ group to relate the two frames via gamma matrices as 
\begin{equation}
\begin{aligned}
   \hE_{\alpha \beta} 
      &= -\tfrac{1}{32}\ii \gamh^{ii'}{}_{\alpha \beta} 
          \big(\hE_{ii'} - \ii \hE'_{ii'}\big) , \\
   \bhE^{\alpha \beta} 
      &= \tfrac{1}{32}\ii \gamh^{ii'}{}^{\alpha \beta} 
          \big(\hE_{ii'} + \ii \hE'_{ii'}\big) ,
\end{aligned}
\end{equation}
such that the components are related by 
\begin{equation}
\label{eq:SU8comp}
\begin{aligned}
   V^{\alpha \beta} &= \ii \gamh_{ii'}{}^{\alpha \beta} 
       \big(V^{ii'} +\ii \tV^{ii'}\big) , \\
   \bV_{\alpha \beta} &= - \ii \gamh_{ii'}{}_{\alpha \beta} 
       \big(V^{ii'} - \ii \tV^{ii'}\big) .  
\end{aligned}
\end{equation}
The orthogonality relations~\eqref{eq:pairSL8} under the symplectic pairing now read 
\begin{equation}
\label{eq:pairSU8}
   \bl{\hE_{\alpha\beta}}{\bar{\hE}^{\gamma\delta}}
      = \tfrac{1}{32}\ii \Phi^2 \delta_{\alpha\beta}^{\gamma\delta} , \qquad
   \bl{\hE_{\alpha\beta}}{\hE_{\gamma\delta}} = 0 , \qquad
   \bl{\bar{\hE}^{\alpha\beta}}{\bar{\hE}^{\gamma\delta}} = 0 .
\end{equation}
For $W\in E^*$ we expand 
\begin{equation}
\label{eq:WSU8}
   W = W_A E^A = \tfrac12\big(
        W^{\alpha\alpha'} E_{\alpha\alpha'} 
        + \bar{W}_{\alpha\alpha'} E^{\alpha\alpha'} \big) ,
\end{equation}
where the components $W^{\alpha\beta}$ and $\bar{W}_{\alpha\beta}$ of the dual vector $W\in E^*$ are defined as in~\eqref{eq:SU8comp}. The contraction~\eqref{eq:contract} then reads 
\begin{equation}
\label{eq:contractSU8}
   V\cdot W = \tfrac{1}{32}\big( V^{\alpha\alpha'}\bar{W}_{\alpha\alpha'}
       + \bar{V}_{\alpha\alpha'} W^{\alpha\alpha'} \big) . 
\end{equation}


\subsection{Torsion}
\label{app:SU8-Dorfman}

In order to calculate the structure constants of the algebra via~\eqref{eq:XY} we need an expression for the projection $T^{(K)}$ of $K=\Dgen-\Dgen'$, the difference in $\SU(8)$ generalised connections, onto the torsion representation. 

The normalisation of the $\SU(8)$ action of $K$ on $\hE_A$ is given by 
\begin{equation}
\begin{aligned}
   K \cdot \hE_{\alpha\beta} 
      &= K^\gamma{}_\alpha\hE_{\gamma\beta} 
         + K^\gamma{}_\beta\hE_{\alpha\gamma} , \\ 
   K \cdot \bhE^{\alpha\beta} 
      &= - K^\alpha{}_\gamma\bhE^{\gamma\beta} 
         - K^\beta{}_\gamma\bhE^{\alpha\gamma} .
\end{aligned}
\end{equation}
where $K$ is a matrix valued-section of $E^*$, satisfying $K=-K^\dag$ and $\tr K=0$, with components, if we expand following~\eqref{eq:WSU8},  
\begin{equation}
   K^\gamma{}_\delta = \tfrac{1}{2}\big
       (K^{\alpha\alpha'\gamma}{}_\delta E_{\alpha\alpha'} 
       + \bar{K}_{\alpha\alpha'}{}^\gamma{}_\delta \bar{E}^{\alpha\alpha'} 
       \big) . 
\end{equation}
Decomposing into $\SU(8)$ representations we have 
\begin{equation}
\begin{aligned}
   K^{\alpha\alpha'\gamma}{}_\delta 
      & \sim \rep{28} + \rep{36} + \rep{420} + \rep{1280} , \\
   \bar{K}_{\alpha\alpha'}{}^\gamma{}_\delta 
      & \sim \rep{\bar{28}} + \rep{\bar{36}} + \rep{\bar{420}} 
         + \rep{\bar{1280}} . 
\end{aligned}
\end{equation}
while the torsion decomposes as 
\begin{equation}
\begin{aligned}
   \rep{56} + \rep{912} &= \rep{28} + \rep{36} + \rep{420}
       + \rep{\bar{28}} + \rep{\bar{36}} + \rep{\bar{420}} , \\
   T &= (T'^{\alpha\beta}, T^{\alpha\beta}, 
       T^{\alpha\beta\gamma}{}_\delta,
       \bar{T}'_{\alpha\beta}, \bar{T}_{\alpha\beta},  
       \bar{T}_{\alpha\beta\gamma}{}^\delta)
\end{aligned}
\end{equation}
where $T'^{\alpha\beta}=-T'^{\beta\alpha}$ is the $\rep{28}$ component and $T^{\alpha\beta}=T^{\beta\alpha}$ is the $\rep{36}$ component, while $T^{\alpha\beta\gamma}{}_\delta=T^{[\alpha\beta\gamma]}{}_\delta$ with  $T^{\alpha\beta\gamma}{}_\gamma=0$ is the $\rep{420}$ component. For the $\SU(8)$ connection $K$, up to normalisation, we have
\begin{equation}
\label{eq:Tcomps}
\begin{aligned}
   T^{(K)\prime\alpha\beta} 
      &\sim K^{\gamma[\alpha\beta]}{}_\gamma , \\
   T^{(K)\alpha\beta} 
      &\sim K^{\gamma(\alpha\beta)}{}_\gamma , \\
   T^{(K)\alpha\beta\gamma}{}_\delta 
      &\sim K^{[\alpha\beta\gamma]}{}_\delta
          - \tfrac{4}{5}\delta^{[\alpha}_\delta
          K^{\beta\gamma\epsilon]}{}_\epsilon .
\end{aligned}
\end{equation}

For the $K$ of interest we will see that only the $\rep{36}$ component is non-zero. However, to calculate~\eqref{eq:XY}, we will need the normalisation of this component compatible with the identification~\eqref{eq:X-spinor} from the algebra~\eqref{eq:alg-SU8}. We first recall that given $V\in E$, the tensor $T(V)$ lies in the adjoint representation of $\E7\times\bbR^+$. Using~\eqref{eq:XT}, we see that the algebra~\eqref{eq:alg-SU8} can be written as 
\begin{equation}
\begin{aligned}
   \Lgen_{\hat{E}_{\alpha\beta}} \hat{E}_{\gamma\delta}
      &= -T^{\hat{\Dgen}}(\hE_{\alpha\beta})\cdot \hE_{\gamma\delta} , & && && 
   \Lgen_{\hat{E}_{\alpha\beta}} \bar{\hat{E}}^{\gamma\delta}
      &= -T^{\hat{\Dgen}}(\hE_{\alpha\beta})\cdot \bhE^{\gamma\delta} , 
\end{aligned}
\end{equation}
where $T^{\hat{\Dgen}}(\hE_{\alpha\beta})$ acts only in the $\SU(8)$ subgroup of $\E7\times\bbR^+$, with 
\begin{equation}
\label{eq:36norm}
   T^{\hat{\Dgen}}(\hE_{\alpha\beta})^\gamma{}_\delta
       = \tfrac{1}{4}\ii\delta^\gamma_{[\alpha}\bar{T}_{\beta]\delta} , 
\end{equation}
and $\bar{T}_{\alpha\beta}=-\bar{X}_{\alpha\beta}
=-R^{-1}\ee^{\ii\omega}\tC_{\alpha\beta}$. 

We now need to calculate the corresponding component of $T^{(K)}$. We start by noting that, with the conventions of section~\ref{app:index}, the generalised Lie derivative~\eqref{eq:Lgen-e7} takes the form
\begin{equation}
\begin{aligned}
   16(\Lgen_V W)^{\alpha \alpha'} 
       = \tfrac{1}{2} &(V^{\gamma\gamma'} \bder_{\gamma\gamma'} 
	+ \bV_{\gamma\gamma'} \der^{\gamma\gamma'} ) W^{\alpha\alpha'} 
        \\ & 
        + 2 (\der^{\alpha \gamma'} \bV_{\gamma\gamma'} ) W^{\gamma \alpha'}
	- 2 (\bder_{\gamma\gamma'} V^{\alpha \gamma'} ) W^{\gamma \alpha'}
	+ \tfrac12  (\bder_{\gamma\gamma'}V^{\gamma\gamma'}) W^{\alpha\alpha'}
        \\ &
        - 3 \big( \der^{[\alpha\alpha'} V^{\beta\beta']} 
           - \tfrac{1}{4!}
           \epsilon^{\alpha\alpha'\beta\beta'\gamma\gamma'\delta\delta'} 
           \bder_{\gamma\gamma'} \bV_{\delta\delta'} \big) 
           \bW_{\beta\beta'} , \\
    16(\Lgen_V W)_{\alpha \alpha'} 
       = \tfrac{1}{2} &(V^{\gamma\gamma'} \bder_{\gamma\gamma'} 
        + \bV_{\gamma\gamma'} \der^{\gamma\gamma'} ) \bW_{\alpha\alpha'} 
        \\ &
        - 2 (\der^{\gamma\gamma'} \bV_{\alpha \gamma'} ) \bW_{\gamma \alpha'}
        + 2 (\bder_{\alpha \gamma'} V^{\gamma\gamma'} ) \bW_{\gamma \alpha'}
        + \tfrac12  (\der^{\gamma\gamma'} \bV_{\gamma\gamma'})
           \bW_{\alpha\alpha'} 
        \\ & 
        -3 \big( \bder_{[\alpha\alpha'} \bV_{\beta\beta']} 
           - \tfrac{1}{4!} 
           \epsilon_{\alpha\alpha'\beta\beta'\gamma\gamma'\delta\delta'} 
           \der^{\gamma\gamma'} V^{\delta\delta'} \big)  W^{\beta\beta'} , 
\end{aligned}
\end{equation}
where again we use the convention that repeated indices with primes are antisymmetrised, so $\alpha\alpha'=[\alpha\alpha']$. If we set $\bar{V}_{\alpha\alpha'}=0$, and keep only $\bar{K}_{\alpha\alpha'}{}^\gamma{}_\delta$, we find
\begin{equation}
\label{eq:TKSU8}
\begin{aligned}
   \big(T^{(K)}(V)\cdot W \big)^{\alpha\alpha'}
      &= \tfrac{1}{16} 
          V^{\gamma\gamma'}\bar{K}_{\gamma\gamma'}{}^\alpha{}_\epsilon
          W^{\epsilon\alpha'}
       - \tfrac{1}{8} \bar{K}_{\gamma\gamma'}{}^\alpha{}_\epsilon 
          V^{\epsilon\gamma'}W^{\gamma\alpha'}
       \\ & \qquad \qquad 
       {} - \tfrac{1}{8} \bar{K}_{\gamma\gamma'}{}^{\gamma'}{}_\epsilon 
          V^{\alpha\epsilon}W^{\gamma\alpha'}
       + \tfrac{1}{16}\bar{K}_{\gamma\gamma'}{}^\gamma{}_\epsilon
          V^{\epsilon\gamma'} W^{\alpha\alpha'} \\
      &= \tfrac{1}{16}V^{\gamma\gamma'}\big( 
          \bar{K}_{\gamma\gamma'}{}^\alpha{}_\epsilon 
            + 2\bar{K}_{\epsilon\gamma}{}^\alpha{}_{\gamma'}
            - 2\bar{K}_{\epsilon\delta}{}^\delta{}_{\gamma'}\delta^\alpha_\gamma
            - \bar{K}_{\delta\gamma}{}^\delta{}_{\gamma'}
                \delta^\alpha_{\epsilon} \big) W^{\epsilon\alpha'} \\
      &= 2 T^{(K)}(V)^\alpha{}_\epsilon W^{\epsilon\alpha'} . 
\end{aligned}
\end{equation}
From the normalisation~\eqref{eq:V-su8-basis} we see that $T^{(K)}(V)=\frac{1}{2}V^{\alpha\beta}T^{(K)}(\hE_{\alpha\beta})$, and from~\eqref{eq:36norm} we see that the $\rep{36}$ component is given by
\begin{equation}
   \bar{T}^{(K)}{}_{\alpha\beta} 
      = - \tfrac{4}{7}\ii T^{(K)}(\bE_{\gamma\alpha})^\gamma{}_{\beta} 
         - \tfrac{4}{7}\ii T^{(K)}(\bE_{\gamma\beta})^\gamma{}_{\alpha} . 
\end{equation}
Comparing with~\eqref{eq:TKSU8} we find
\begin{equation}
\label{eq:TK36}
   \bar{T}^{(K)}{}_{\alpha\beta} 
      = -\tfrac{1}{2} \ii \bar{K}_{\delta(\alpha}{}^\delta{}_{\beta)} . 
\end{equation}


\subsection{$\SU(4)$ decomoposiiton of $\{\hF_A\}$}
\label{app:su4-decomp}

We start by calculating the form of the singlet $\hF_{12}$ in the decomposition~\eqref{eq:F-decomp} of $\{\hF_A\}$. Fixing the $\SU(4)$ singlet spinors $\eta_1$ and $\eta_2$ by\footnote{The minus sign before $\gamh^{78}$ here is necessary as we are in the negative chirality representation~\eqref{eq:-ve-chirality}.}
\begin{equation}
\label{eq:eta-spinor-def}
   \gamh^{12}\eta_2 = \gamh^{34}\eta_2 = \gamh^{56}\eta_2
      = -\gamh^{78}\eta_2 = \eta_1
\end{equation}
we note that, using~\eqref{eq:SU8-basis},
\begin{equation}
   \hF_{12} = \eta_1^\alpha\eta_2^\beta \hF_{\alpha\beta}
      = - \tfrac{1}{32}\ii 
         \big(\eta_1^\alpha\gamh^{ij}_{\alpha\beta}\eta_2^\beta\big)
         \big( \hF_{ij} - \ii\hF'_{ij} \big) 
      = - \tfrac{1}{32}  \ii \Omega^{ij} 
         \big( \hF_{ij} - \ii\hF'_{ij} \big) .  
\end{equation}
Let us choose a gauge where the six-form potential is given by 
\begin{equation}
   \tA_{(6)} = -\tfrac12 R^6 \dd \psi \wedge A \wedge \omegaB^2
\end{equation}
where $A$ is the Hopf fibration connection~\eqref{eq:A-def}. It is easy to check this leads to the desired flux
\begin{equation}
   \tF_{(7)} = \dd \tA_{(6)} = R^6 \dd \psi \wedge \omegaB^3 
       = 6R^{-1} \vol_{S^7} . 
\end{equation}
Using
\begin{equation}
   i_{\der_\psi}\tA_6 = -\tfrac12 R^6 A \wedge \omegaB^2 , \qquad
   j\tA \wedge \omega_B = RA \otimes \vol_g ,
\end{equation}
and the definitions~\eqref{eq:E7para} and relations~\eqref{eq:singlets} we find 
\begin{equation}
\begin{aligned}
   \tfrac{1}{2}\Omega_{ij} \hF^{ij} 
     &= R^{-1}\der_\psi + R^5\lambda\wedge\tfrac{1}{2}\omegaB^2 
         - R^6A\wedge\tfrac{1}{2}\omegaB^2 \\
     &= R^{-1}\der_\psi + R^5\dd\psi\wedge\tfrac{1}{2}\omegaB^2 , \\
   \tfrac{1}{2}\Omega_{ij} \hF^{\prime ij} 
     &= R^2\omegaB + R\lambda\otimes \vol_g - RA\otimes \vol_g \\
     &= R^2\omegaB + R\dd\psi\otimes \vol_g  .
\end{aligned}
\end{equation}
and hence 
\begin{equation}
\label{eq:F12}
   \hF_{12} = - \tfrac{1}{16}\ii\big( R^{-1}\der_\psi 
          + R^5\dd\psi\wedge\tfrac{1}{2}\omegaB^2 \big) 
      - \tfrac{1}{16}\big( R^2\omegaB + R\dd\psi\otimes \vol_g \big). 
\end{equation}

Rather than calculate the other $\SU(4)$ components in~\eqref{eq:F-decomp} explicitly, for the calculation in section~\ref{sec:frame}, we only need to know the combinations of $\der_\psi$, $\omegaB$, $\dd\psi\wedge\omegaB$ and $\dd\psi\otimes\vol_g$ that appear in $\hF_{\talpha\tbeta}$, $\hF_{\talpha 1}$ and $\hF_{\talpha 2}$. The easiest way to calculate this is using the symplectic invariant of $\E7$.
Recall that the frame $\{\hF_A\}$ satisfies the orthogonality relations~\eqref{eq:pairSU8} with $\Phi^2=\ee^{2\Delta}\vol_g$. Let us focus on the subspace of $E$ spanned by $\der_\psi$, $\omegaB$, $\dd\psi\wedge\omegaB$ and $\dd\psi\otimes\vol_g$. Using the fact the symplectic pairing of $\hF_{\talpha\tbeta}$, $\hF_{\talpha 1}$ and $\hF_{\talpha 2}$ with both $\hF_{12}$ and $\bar{\hF}^{12}$ must vanish, we find 
\begin{equation}
\label{eq:nonF12}
   \left.\begin{matrix}
      \hF_{\talpha\tbeta} \\ \hF_{\talpha 1} \\ \hF_{\talpha 2}
   \end{matrix}\right\}
   = (\dots) \big( R^{-1}\der_\psi 
          - \tfrac{1}{6}R^5\dd\psi\wedge\omegaB^2 \big) 
     + (\dots) \big( \tfrac{1}{3}R\omegaB - R\dd\psi\otimes \vol_g \big) 
     + \dots ,  
\end{equation}
since only these combinations have vanishing symplectic paring with $\hF_{12}$ and $\bar{\hF}^{12}$. 

Finally we note that, since $\dd\omegaB=\dd(\dd\psi\wedge\omegaB^2)=0$, we have from~\eqref{eq:F12} and~\eqref{eq:Lgen-e7}
\begin{equation}
   \Lgen_{\hF_{12}} = -\tfrac{1}{16}\ii  R^{-1} \mathcal{L}_{\der_\psi} . 
\end{equation}
From the algebra~\eqref{eq:alg-SU8}, we have 
\begin{equation}
   \Lgen_{\hF_{12}} \hF_{\talpha\tbeta} = 0 , \qquad
   \Lgen_{\hF_{12}} \hF_{\talpha1} 
       = \tfrac{1}{8}\ii R^{-1} \hF_{\talpha2} , \qquad 
   \Lgen_{\hF_{12}} \hF_{\talpha2}
       = -\tfrac{1}{8}\ii R^{-1} \hF_{\talpha1} , 
\end{equation}
Hence we can conclude that, in the coordinates~\eqref{eq:CP3-coord}, $\hF_{\talpha\tbeta}$ is independent of $\psi$ while for  $\hF_{\talpha\pm}=\hF_{\talpha1}\pm\ii\hF_{\talpha2}$ we have $\mathcal{L}_{\der_\psi}\hF_{\talpha\pm}=\pm2\ii \hF_{\talpha\pm}$ and hence have the dependence
\begin{equation}
   \hF_{\talpha\pm} = \ee^{\pm2\ii\psi} \big[ \dots \big] . 
\end{equation}
%


\subsection{Calculation of $T^{(K)}$ }
\label{app:TK-calc}

We first need to satisfy the condition~\eqref{eq:S7-f-properties} on $f$. Given the form of~\eqref{eq:F12} and~\eqref{eq:nonF12} and the identifications~\eqref{eq:gen-coords}, we require
\begin{equation}
\begin{aligned}
   \left( \der_{\psi_0}-\tfrac{1}{3}\der_{\psi_4}\right)f &= 0 , \\
   \left( \tfrac{1}{3}\der_{\psi_2}-\der_{\psi_6}\right)f &= 0 , \\
   -\tfrac{1}{16}\ii\left( \der_{\psi_0}+\der_{\psi_4}\right)f 
   -\tfrac{1}{16}\left( \der_{\psi_2}+\der_{\psi_6}\right)f 
         &= c , 
\end{aligned}
\end{equation}
which has the solution
\begin{equation}
   f = a(\psi_0 + 3 \psi_4) + b(\psi_2 + \tfrac13 \psi_6) , \qquad
   c = -\tfrac{1}{12} b- \tfrac14 \ii  a . 
\end{equation}
We then have, since $K^\alpha{}_\beta=-(\dd f)\lambda^\alpha{}_\beta$ and $\hF_{\alpha\beta} (f) = c R^{-1} \bar\kappa_{\alpha\beta}$, 
\begin{equation}
   K(\hF_{\alpha\beta})^\gamma{}_\delta
      = -cR^{-1}\bar{\kappa}_{\alpha\beta}\lambda^\gamma{}_\delta . 
\end{equation}
where $\lambda$ is given by~\eqref{eq:lambda-def} and $\kappa_{\alpha\beta}= \eta_1^\alpha\eta_2^\beta-\eta_2^\alpha\eta_1^\beta$. From~\eqref{eq:Tcomps}, we see that the $\rep{420}$ and $\rep{28}$ parts of $T^{(K)}$ vanish. For the $\rep{36}$ component, we first note that the normalisations~\eqref{eq:V-su8-basis} and~\eqref{eq:contractSU8} lead to $K(\hF_{\alpha\beta})^\gamma{}_\delta = \tfrac{1}{16} \bar{K}_{\alpha\beta}{}^\gamma{}_{\delta}$, and then from~\eqref{eq:TK36} we have
\begin{equation}
   \bar{T}^{(K)}{}_{\alpha\beta} = 8\ii cR^{-1} \bar{L}_{\alpha\beta}
      = 2  R^{-1}(a-\tfrac{1}{3}\ii b) \bar{L}_{\alpha\beta} . 
\end{equation}
%


\section{Type IIA and charges}
\label{app:charges}

The reduction of $S^7/\bbZ_p$ on the Hopf fibre to IIA is presented in~\cite{ABJM}. We use the conventions of~\cite{csw1}, though here we dualise the ``democratic" RR fluxes $F_{(6)}$ and $F_{(8)}$ in the equations of motion, thus keeping only $F_{(2)}$ and $F_{(4)}$. Equivalently, one can merely start from the action
\begin{equation}
\label{eq:IIA-action}
   S = \frac{1}{2 \kappa^2} \int \Big[ \sqrt{-g} \Big(
       \ee^{-2\phi} ( \mathcal{R} + 4 (\der \phi)^2 - \tfrac{1}{12} H^2 )
          - \tfrac12 F^2_{(2)} - \tfrac12 F^2_{(4)} \Big)
          - \tfrac14 B_{(2)} \wedge F_{(4)} \wedge F_{(4)} \Big] .
\end{equation}

The type IIA string frame metric is related to the eleven-dimensional on $\AdS_4\times S^7/\bbZ_p$ in M theory units by
\begin{equation}
\label{eq:IIA-metric}
   g_{\IIA} = \frac{R}{p} \left( \frac14 R^2 \dd s^2_{\AdS_4} 
      + R^2 \dd s^2_{\CP^3} \right) ,
\end{equation}
where $\dd s^2_{\CP^3}$ is the standard metric~\eqref{eq:CP3-metric} and flux quantisation requires that
\begin{equation}
   R^6 = 32\pi^2 pq ,
\end{equation}
for some integer $q$. The remaining non-zero IIA fields are given by
\begin{equation}
   \ee^{2\phi} = \left( \frac{R}{p} \right)^3 , 
   \qquad
   F_2 = 2p \omega , 
   \qquad
   F_4 = \frac{3 R^3}{8} \vol_{\AdS_4} .
\end{equation}
Noting that the unit $\AdS_4$ and $\CP^3$ spaces are Einstein with the respective Ricci tensors
\begin{equation}
   \mathcal{R}_{\AdS_4} = - 3 g_{\AdS_4} , 
   \qquad
   \mathcal{R}_{\CP^3} =  + 8 g_{\CP^3} ,
\end{equation}
we see that these satisfy the ten-dimensional equations of motion following from the action~\eqref{eq:IIA-action}.

We now calculate the masses of $D$-branes wrapping the non-trivial cycles of $\CP^3$ following~\cite{Aharony:2009fc,Gutierrez:2010bb}. The mass of a D$n$-brane in type IIA units is given by (see e.g.~\cite{Polchinski})
\begin{equation}
   M_{n,\IIA} = \tau_n \Vol(C) , \qquad
   \tau_n = g^{-1}_s (2\pi)^{-n} , 
\end{equation}
where $g_s = \ee^{\phi}$ is the string coupling, $\Vol(C)$ is the volume of the brane wrapping $C$ and we have set $\alpha' = 1$. We are interested in the D$n$-branes wrapping the $\CP^n$ cycles of $\CP^3$. We wish to define circles spanned by extra coordinates, whose radii are such that the first KK modes around these dual circles give precisely the masses of the wrapped branes. We therefore set
\begin{equation}
   R_{n,\IIA} = [ M_{n,\IIA} ]^{-1} .
\end{equation}
From~\eqref{eq:IIA-metric}, there is an overall factor which gives the conversion between string units and M theory units
\begin{equation}
   (\text{Length}_{M}) 
      = \left(\frac{p}{R}\right)^{1/2} (\text{Length}_{\IIA}) ,
\end{equation}
so we define
\begin{equation}
   R = \left(\frac{p}{R}\right)^{1/2} R_{\IIA} 
   \qquad \text{and} \qquad 
   R_{n,M} = \left(\frac{p}{R}\right)^{1/2} R_{n,\IIA} .
\end{equation}
Using the standard normalisations as in section~\ref{app:hopf}, we have the volume forms $\vol_{\CP^n} = \tfrac{1}{n!} \omegaB^n$ and $\int_{\CP^n}\omegaB^n= \pi^n$. Hence, in IIA units we have the volumes  
\begin{equation}
   \Vol(\CP^n) = \tfrac{1}{n!} \pi^n R^{2n}_{\IIA} .
\end{equation}
The masses of the wrapped D-branes in IIA units are thus
\begin{equation}
   M_{n,\IIA} 
   = \frac{p}{R_{\IIA}} \left( \frac{R^2_{\IIA}}{2} \right)^n .
\end{equation}
All this results in the radii for the dual circles (in M theory units)
\begin{equation}
\begin{aligned}
   R_{0,M} = \frac{R}{p} \qquad R_{2,M} &= \beta \frac{R}{p}  , \\
   R_{4,M} = \frac{R}{q} \qquad R_{6,M} &= 3\beta \frac{R}{q} , 
\end{aligned}
\end{equation}
where $\beta = \sqrt{\frac{p}{2q}}$.




\end{document}